\documentclass[aps,pra,superscriptaddress,twocolumn]{revtex4}
\usepackage{extarrows}
\usepackage{slashed}
\usepackage{amsmath,amsfonts,amssymb,graphics,graphicx,epsfig,color,times,indentfirst,layout}
\usepackage{lipsum}
\usepackage{natbib}

\begin{document}

\title{Entanglement negativity after a local quantum quench in conformal field theories}

\date{\today}

\author{ Xueda Wen}
\affiliation{Department of Physics, University of Illinois at Urbana-Champaign, Urbana, IL 61801, USA}
\author{ Po-Yao Chang}
\affiliation{Department of Physics, University of Illinois at Urbana-Champaign, Urbana, IL 61801, USA}
\author{ Shinsei Ryu}
\affiliation{Department of Physics, University of Illinois at Urbana-Champaign, Urbana, IL 61801, USA}

\begin{abstract}

We study the time evolution of the entanglement negativity after a local quantum quench in
(1+1)-dimensional conformal field theories (CFTs),
which we introduce by suddenly joining two initially decoupled CFTs at their endpoints.
We calculate the negativity evolution for both adjacent intervals and disjoint intervals explicitly.
For two adjacent intervals, the entanglement negativity grows
logarithmically in time right after the quench.
After developing a plateau-like feature, the entanglement negativity drops to the ground-state value.
For the case of two spatially separated intervals, a light-cone behavior is observed in the negativity evolution;
in addition, a long-range entanglement, which is independent of the distance between two intervals,  can be created.
Our results agree with the heuristic picture that quasiparticles, which carry entanglement,  are emitted from the
joining point and propagate freely through the system. Our analytical results are confirmed by numerical calculations based on a critical harmonic chain.
\end{abstract}
\maketitle


\section{Introduction}

\subsection{Introduction}
Recently, it has been recognized that quantum entanglement provides us a powerful tool to study quantum properties of many-body systems in condensed matter physics~\cite{Kitaev2006,LevinWen2006,Calabrese0905,Eisert0808}.
When the system is prepared in a \emph{pure} state $|\Psi\rangle$, a good quantity that describes the bipartite entanglement
is the von Neumann entropy, which is defined as
\begin{equation}
S_A=-\text{Tr}\rho_A\ln\rho_A
\end{equation}
where $\rho_A=\text{Tr}_B\rho$ is the reduced density matrix of subsystem $A$, with $\rho=|\Psi\rangle\langle\Psi|$.
An alternative measure of bipartite entanglement in pure states is the Renyi entropy
\begin{equation}\label{RenyiS}
S_A^{(n)}=\frac{1}{1-n}\ln \text{Tr}\rho_A^n.
\end{equation}
These entanglement measures have proved to be of great use in characterizing quantum entanglement of many-body states.

However, for a mixed state,
neither the von Neumann entropy nor the Renyi entropy is a good measure of entanglement
since quantum and classical correlations are not clearly separated in these measures.
Now suppose we are interested in the entanglement between two subsystems $A_1$ and $A_2$,
which are not necessarily complementary to each other and, are embedded in a larger system,
the union $A_1\cup A_2$ cannot be described by a pure state
after integrating out degrees of freedom in the complement of $A_1 \cup A_2$.
In this case, we need to search for other quantities that may characterize quantum entanglement for a general mixed state.
Among different proposals
\cite{Plenio2007,Plenio1999},
a computable measurement of entanglement, the logarithmic negativity
\cite{Vidal2002}, turns out to be very useful and practical.
In particular, it is proved that the logarithmic negativity is a proper entanglement monotone in Ref.\ \cite{Plenio2005}.
Following Ref.\ \cite{Vidal2002}, the negativity can be obtained by first taking a partial transposition.
To be more precise, given a density matrix $\rho_{A_1\cup A_2}$
which describes a bipartite mixed state in a Hilbert space $\mathcal{H}_{A_1\cup A_2}=\mathcal{H}_{A_1}\otimes\mathcal{H}_{A_2}$,
the partial transposition with respect to $A_2$'s degrees of freedom is defined as
\begin{equation}
\langle e_i^{(1)}e_j^{(2)}|\rho_{A_1\cup A_2}^{T_2}|e_k^{(1)}e_l^{(2)}\rangle=
\langle e_i^{(1)}e_l^{(2)}|\rho_{A_1\cup A_2}|e_k^{(1)}e_j^{(2)}\rangle,
\end{equation}
where $|e_i^{(1)}\rangle$ and $|e_j^{(2)}\rangle$ are arbitrary bases
in
$\mathcal{H}_{A_1}$ and $\mathcal{H}_{A_2}$, respectively.
Then the logarithmic negativity is defined as
\begin{equation}
\mathcal{E}_{A_1,A_2}:=\ln||\rho_{A_1\cup A_2}^{T_2}||=\ln\text{Tr}|\rho_{A_1\cup A_2}^{T_2}|,
\end{equation}
where $T_2$ indicates the partial transposition with respect to $A_2$,
and the trace norm $||\rho_{A_1\cup A_2}^{T_2}||$
is defined as
the sum of all the absolute values of
the eigenvalues of $\rho_{A_1\cup A_2}^{T_2}$.
Recently, the logarithmic negativity has been extensively used to study various many-body systems,
including one-dimensional harmonic chains
\cite{Audenaert2002,Marcovitch2009},
quantum spin chains
\cite{Wichterich2009,Wichterich2010, Bayat2010a,Bayat2010b,Bayat2010c,Santos2011a,Santos2011b},
free fermion systems
\cite{Chang2015},
and topologically ordered systems
\cite{Vidal2013,Castelnovo2013}.
In particular, the universal features of the entanglement negativity in one-dimensional critical systems
have been understood by developing a conformal field theory (CFT) approach
\cite{Calabrese2012a,Calabrese2012b}.
Later on, the comparison of CFT results and numerical calculations of one-dimensional critical systems were studied
in a series of works
\cite{Calabrese1302,Alba1302, Alba2014}.

Although many works have been done on the entanglement negativity,  there is less understanding on the non-equilibrium properties of
the entanglement negativity.
Most recently, time evolution of the logarithmic negativity after a global quench was studied with CFT approach \cite{Coser}.
In Ref.\ \cite{Viktor},
the negativity evolution for two adjacent intervals after a local quench was numerically studied in a harmonic chain.
However, a thorough study of the negativity evolution after a local quantum quench is still lacking,
and it is appealing to unveil the universal features of the dynamical behavior of
the entanglement negativity propagation.

\begin{figure}
\includegraphics[scale=0.45]{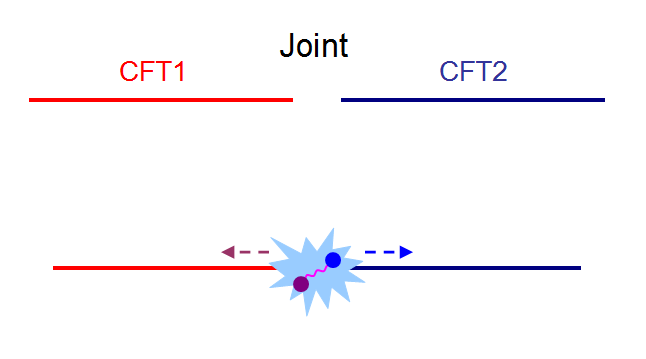}
\caption{
Setup for a local quantum quench.
Two separate CFTs defined on two semi-infinite lines are joined together at their endpoints.
Then quasiparticles, which may be viewed as entangled pairs,
are generated at the jointing point, and propagate freely through the system. The entanglement negativity between two intervals which are far from each other may be built with the help of these propagating entangled pairs.}\label{Setup}
\end{figure}

In this paper, our motivation is to study the time evolution of the entanglement negativity after a local quantum quench analytically.
For simplicity,
we consider a (1+1)-dimensional critical system,
which is physically cut into two parts that are prepared in their own ground states.
Then at time $t=0$, we join the two parts together at their endpoints,
and
study
the time-evolution of the entanglement negativity afterwards.
As shown in Fig.\ \ref{Setup}, once the two CFTs are joined at the endpoints, the interaction between them is introduced simultaneously,
which generates quasiparticles (excitations) at the jointing point.
These quasiparticles may be viewed as entangled pairs \cite{Tadashi1302, Tadashi1401,Tadashi1403,Calabrese2006PRL,Calabrese2007quench}
which carry entanglement information. When the entangled pairs arrive at two intervals separately, the entanglement negativity can be built immediately.
Because the (1+1) dimensional critical system is Lorentz invariant at the low energy limit, we can utilize the power of conformal field theory and understand the universal feature of this dynamical phenomenon.

The rest of the paper is organized as follows.
In part B of this section, we give a brief review of the  path integral representation of the entanglement negativity,
and then introduce the CFT setup for a local quantum quench in part C.
In Section II, by using CFT approach, we compute the time evolution of the entanglement negativity for two adjacent intervals in part A,
and two disjoint intervals in part B.
We consider both cases where the two intervals are symmetrically and asymmetrically located.
In section III, we describe the numerical method of calculating the entanglement negativity for a harmonic chain,
based on which we study the local quench of the entanglement negativity.
Then we compare the numerical results with the CFT results. In section IV, we conclude our work and list some interesting future problems to be studied.

\subsection{Entanglement negativity in quantum field theory}

A detailed description of path integral representation of the entanglement negativity can be found in Ref.\ \cite{Calabrese2012b}.
For the completeness of this paper, we give a brief review here.

First, as discussed in Ref.\ \cite{Calabrese2012b}, by using a replica trick,
one can relate the entanglement negativity with the integer powers of $\rho_{A_1\cup A_2}^{T_2}$ as
\begin{equation}\label{NegativityDef}
\mathcal{E}_{A_1,A_2}=\lim_{n_e\to 1}\ln\text{Tr}\left(\rho_{A_1\cup A_2}^{T_2}\right)^{n_e},
\end{equation}
where $n_e$ is an even integer, and the density matrix $\rho$ may be expressed as a (Euclidean) path integral
in the imaginary time interval $\tau\in[0,\beta]$:
\begin{equation}
\begin{split}
\rho=\frac{1}{Z}\int\left[d\phi(x,\tau)\right]\prod_x\delta\left(\phi(x,0)-\phi'(x')\right)\\
\times
\prod_{x}\delta\left(\phi(x,\beta)-\phi^{\prime\prime}(x^{\prime\prime})\right)e^{-S_E},
\end{split}
\end{equation}
where the rows and columns of the density matrix are labeled by the fields $\{\phi(x,\tau)\}$ at $\tau=0$ and
$\beta$ respectively, with $\beta$ being the inverse temperature,
$S_E$ is the Euclidean action and $Z=\text{Tr}\, e^{-\beta H}$ is the partition function.
Now we consider subsystems $A_1$ and $A_2$ located in intervals $[u_1,v_1]$ and $[u_2,v_2]$, respectively.
Then the reduced density matrix $\rho_{A_1\cup A_2}$ may be obtained by sewing together all the points along
edges $\tau=0$ and $\tau=\beta$ except the points in $A_1\cup A_2$.
That is, we leave two open cuts at $[u_1,v_1]$ and $[u_2,v_2]$ along $\tau=0$.

\begin{figure*}
\includegraphics[scale=0.51]{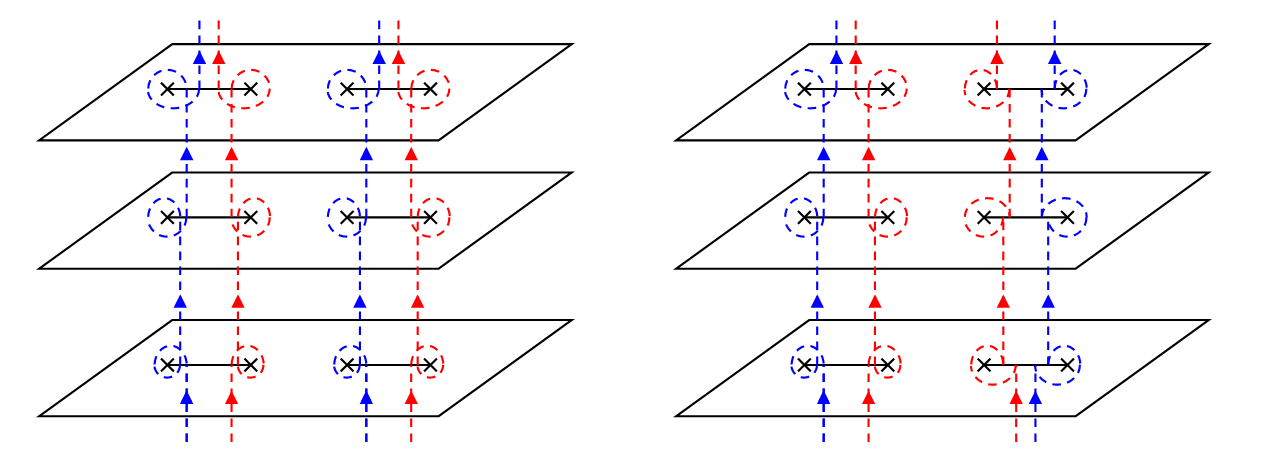}
\caption{Path integral representation of (a) $\text{Tr}(\rho_A)^n$ and (b) $\text{Tr}(\rho_A^{T_2})^n$ for two disjoint intervals.
}\label{PathIntegral}
\end{figure*}

Next, before we compute $\text{Tr}\, (\rho_{A_1\cup A_2}^{T_2})^{n_e}$,
it is beneficial to see how to calculate $\text{Tr}\left(\rho_{A_1\cup A_2}\right)^{n}$ first.
In order to calculate $\text{Tr}\left(\rho_{A_1\cup A_2}\right)^{n}$, we consider $n$ copies of the cut plane, and then sew together the cut $[u_i,v_i]_{\tau=0^-}^{j}$
with the cut $[u_i,v_i]_{\tau=0^+}^{j+1}$ for $i=1,2$ and all the copies $j=1,\cdots,n$. Note that for $j=n$, we sew together the cut $[u_i,v_i]_{\tau=0^-}^{j=n}$
with the cut $[u_i,v_i]_{\tau=0^+}^{j=1}$. In this way, we define a $n$-sheeted Riemann surface $\mathcal{R}_n$.
The trace of $\left(\rho_{A_1\cup A_2}\right)^{n}$ is then given by
\begin{equation}\label{traceNcopy}
\text{Tr}\left(\rho_{A_1\cup A_2}\right)^{n}=\frac{Z_{\mathcal{R}_n}}{Z^n},
\end{equation}
where $Z_{\mathcal{R}_n}$ is the partition function for the orbifold CFT on $\mathcal{R}_n$.
Rather than dealing with the fields on a nontrivial manifold, it is found more convenient to work on a single complex plane.
It turns out Eq.\ (\ref{traceNcopy}) can be expressed in terms of local twisted fields defined at $(u_i,0)$ and $(v_i,0)$ on the complex plane as follows
\begin{equation}
\text{Tr}\left(\rho_{A_1\cup A_2}\right)^{n}=\left\langle
\mathcal{T}_n(u_1)\bar{\mathcal{T}}_n(v_1)\mathcal{T}_n(u_2)\bar{\mathcal{T}}_n(v_2)
\right\rangle.
\end{equation}
Intuitively, the effect of twist fields $\mathcal{T}_n$ and $\bar{\mathcal{T}}_n$ is shown in Fig.\ \ref{PathIntegral}.
Winding anticlockwise (clockwise) around the twist field $\mathcal{T}_n$ ($\bar{\mathcal{T}}_n$), once the branch cut is crossed, one will go from layer $j$ to layer $j+1$.

With the introduction of twist fields,
the expression of $\text{Tr}\, (\rho_{A_1\cup A_2}^{T_2})^{n}$ is very straightforward.
As discussed in Refs.\ \cite{Calabrese2012a,Calabrese2012b},
the effect of partial transposition with respect to $A_2$ is equivalent to changing the two twist operators $\mathcal{T}_n(u_2)$ and $\bar{\mathcal{T}}_n(v_2)$. Then one has
\begin{equation}\label{Disjoint}
\text{Tr}\left(\rho_{A_1\cup A_2}^{T_2}\right)^{n}=\left\langle
\mathcal{T}_n(u_1)\bar{\mathcal{T}}_n(v_1)\bar{\mathcal{T}}_n(u_2)\mathcal{T}_n(v_2)
\right\rangle.
\end{equation}
If the two intervals $[u_1,v_1]$ and $[u_2,v_2]$ are adjacent to each other, we simply set $u_2\to v_1$, and then Eq.\ (\ref{Disjoint}) can be written as
\begin{equation}\label{Adjacent}
\text{Tr}\left(\rho_{A_1\cup A_2}^{T_2}\right)^{n}=\left\langle
\mathcal{T}_n(u_1)\bar{\mathcal{T}}^2_n(u_2)\mathcal{T}_n(v_2)
\right\rangle.
\end{equation}
Therefore, from Eqs.\ (\ref{NegativityDef}), (\ref{Disjoint}) and (\ref{Adjacent}),
it is found that the computation of the entanglement negativity reduces to the computation of expectation values of twist fields in a complex plane.

\subsection{CFT approach to a local quench}

Before we study the CFT approach to a local quantum quench, it is beneficial to comment on the difference between local quenches and global quenches.
Local quenches are more complicated than global quenches because they are inhomogeneous.
For global quenches, we change the parameters of a translational invariant Hamiltonian globally, and therefore the system before and after global quenches are always translational invariant.
In this case, as discussed in Refs.\cite{Calabrese2006PRL,Calabrese2007quench}, the initial state can flow to a conformal invariant boundary state under renormalization group (RG).
For local quenches, however, before we join the two decoupled CFTs together, the total system is apparently not translational invariant.
Therefore, the initial state cannot flow to a conformal invariant boundary state. In addition, for global quenches, quasiparticle excitations are emitted from everywhere in the bulk of CFT; for local quenches, quasiparticle excitations are emitted only from the point where two CFT are joined together.

\begin{figure}[htp]
\includegraphics[scale=0.42]{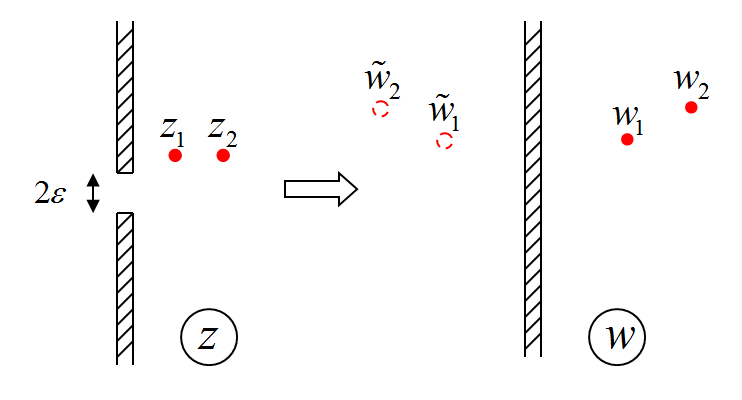}
\caption{Illustration of the conformal mapping in Eq.(\ref{conformalMap}), based on which the
$z$-plane is mapped to a right half plane (RHP) with $\text{Re}\, w>0$. For later use, we also label $\widetilde{w}_i=-\bar{w}_i$, which is the image of $w_i$. }\label{cM}
\end{figure}

The time dependent density matrix can be written as $\rho(t)=|\phi(x,t)\rangle\langle \phi(x,t)|$, where $|\phi(x,t)\rangle=e^{-iHt}|\phi_0(x)\rangle$. In path integral representation, one has
\begin{equation}
\begin{split}
&\langle \phi^{\prime\prime}(x^{\prime\prime})|\rho(t)|\phi'(x')\rangle\\
=&\frac{1}{Z}\langle \phi^{\prime\prime}(x^{\prime\prime})|e^{-iHt-\epsilon H}|\phi_0(x)\rangle\langle\phi_0(x)|e^{+iHt-\epsilon H}|\phi'(x')\rangle,
\nonumber
\end{split}
\end{equation}
where the factor $e^{-\epsilon H}$ is introduced to damp out high-energy modes and make the path integral absolutely convergent.
If the CFT arises as the low energy limit of a lattice model, then $\epsilon$ may be viewed as the lattice spacing.
In the study of global quenches\cite{Calabrese2006PRL,Calabrese2007quench,Calabrese2007},
$|\phi_0(x)\rangle$ may be considered as a conformal invariant boundary state under the  RG.
For local quenches, as we discussed above, the initial state cannot flow to a conformal invariant boundary state under the RG.
In this case, one may introduce boundary condition changing operators, as utilized in Refs.\ \cite{Affleck,Fradkin2009}.
In this work, however, we will follow the method proposed by Calabrese and Cardy
\cite{Calabrese2007}.
As shown in Fig.\ref{cM},
the density matrix can be expressed in terms of the path integral on a modified word-sheet,
where the physical cut corresponds to two `walls' with one extending from $\tau=-\infty$ to $-\epsilon$ and the other extending from $\tau=+\epsilon$ to $+\infty$ in a complex $z$-plane.
No energy nor momentum can flow through the two `walls', and therefore conformal boundary conditions are imposed on the wall
(As will be shown later, the concrete boundary condition does not affect the universal result we consider.).
For convenience of calculation, we map the $z$-plane to a right half plane (RHP) with $\text{Re}\, w>0$ by using the conformal mapping
\begin{equation}\label{conformalMap}
w=\frac{z}{\epsilon}+\sqrt{\left(\frac{z}{\epsilon}\right)^2+1}.
\end{equation}
Then the local quench problem is reduced to the calculation of correlation functions of twist fields in the RHP
\cite{Calabrese2007},
which we will study in detail in the next section.

\section{Entanglement negativity after a local quench: Conformal field theory approach}

In this section, we calculate the time evolution of the entanglement negativity after a local quench in conformal field theories.
We will consider adjacent intervals in part A and disjoint intervals in part B, respectively.

\subsection{Two adjacent intervals}
\begin{figure}
\includegraphics[scale=0.42]{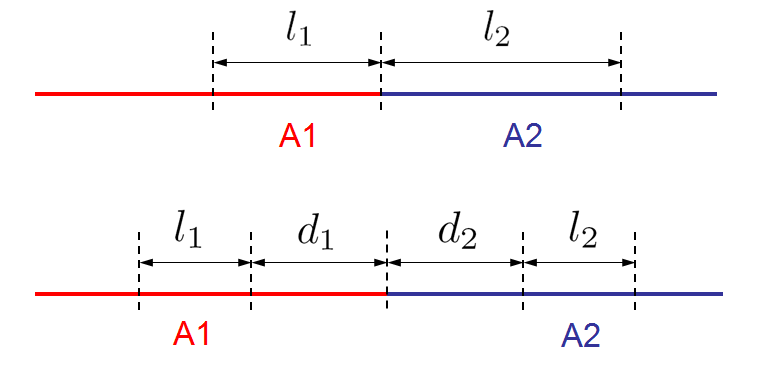}
\caption{Configurations of two intervals considered in this work: two adjacent intervals (up) and two disjoint intervals (bottom). }\label{dL}
\end{figure}

\subsubsection{Semi-infinite intervals}
\begin{figure*}
\includegraphics[width=6.25in]{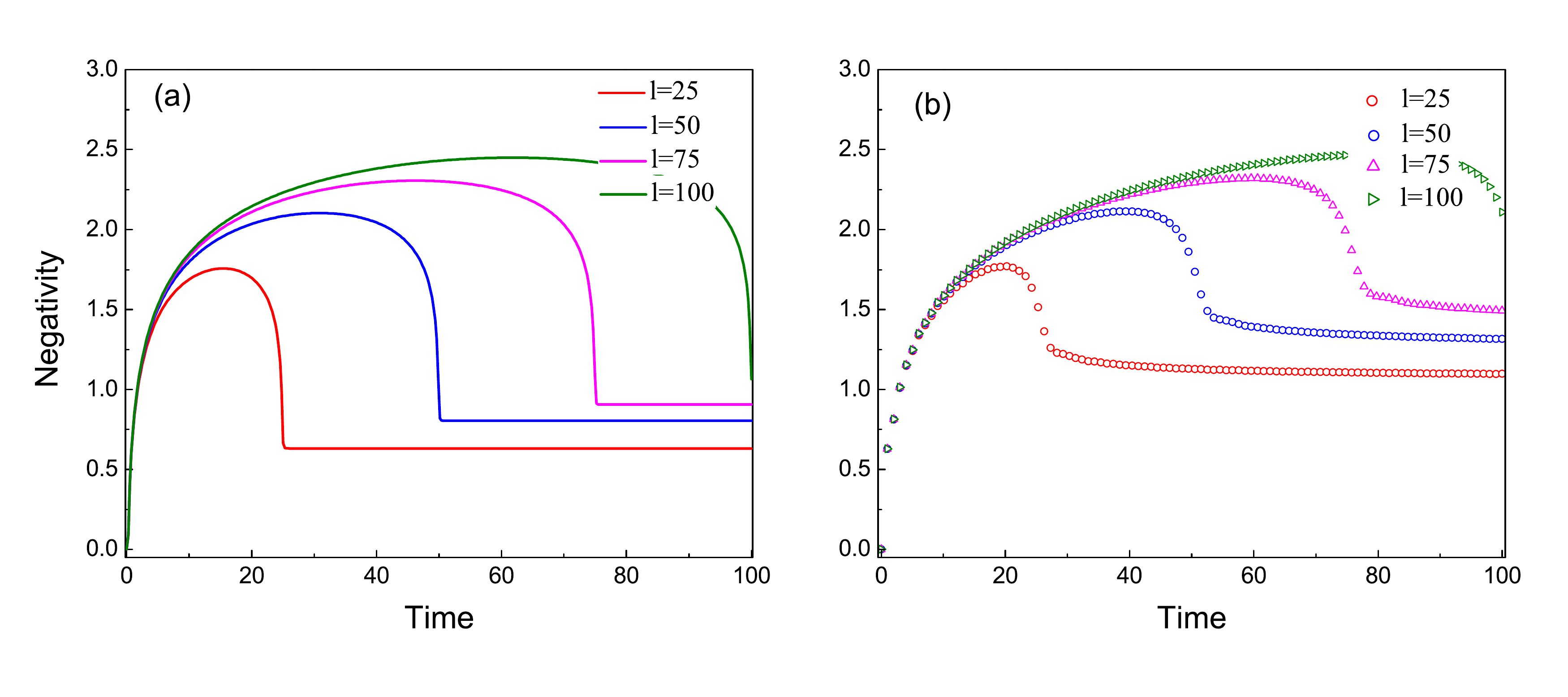}
\caption{Entanglement negativity $\mathcal{E}$ for two symmetric adjacent intervals as a function of time.
Here we choose the central charge $c=1$, $\epsilon=0.1$, $l=25$, $50$, $75$ and $100$, respectively. Shown in (a) is the CFT result, and (b) is the numerical calculation based on a critical harmonic chain. }\label{SymmetricAdjacentFig}
\end{figure*}

As a warm up, we consider the simplest case, i.e.,
the total system is bipartitioned into two semi-infinite parts $A_1$ and $A_2$.
In this case, $\rho_{A_1\cup A_2}$ is pure, and the logarithmic negativity is the same as the Renyi entropy with $n=1/2$
\cite{Calabrese2012a,Calabrese2012b}.
This case was also studied in Ref.\ \cite{Viktor}.

For two adjacent semi-infinite intervals, we only need to consider a single twist field
$\mathcal{T}^2_n(z_1)$ in $z$-plane, which is inserted at
\begin{equation}
z_1=l+i\tau.
\end{equation}
By choosing the insertion position at the origin $l=0$, i.e., $A_1\in (-\infty,0]$ and $A_2\in(0,+\infty)$, one simply has $z_1=i\tau$.
The expectation value of $\mathcal{T}^2_n(z_1)$ can be expressed as \cite{Calabrese2007}
\begin{equation}
\left\langle \mathcal{T}^2_n(z_1)\right\rangle
=\tilde{c}_n
\left(\left|\frac{dw}{dz}\right|_{z_1}\frac{a}{2\text{Re}(w_1)}\right)^{\Delta^{(2)}_{n}},
\end{equation}
where $\tilde{c}_n$ is a nonuniversal constant which depends on the particular boundary CFT,
$a$ is an UV cutoff (\emph{e.g.}, the lattice spacing in a harmonic chain),
and $\Delta^{(2)}_{n}$ is the scaling dimension of $\mathcal{T}^2_n(z)$. By using the conformal map in Eq.\ (\ref{conformalMap}), one has
\begin{equation}
w_1=i\frac{\tau}{\epsilon}+\frac{1}{\epsilon}\sqrt{\epsilon^2-\tau^2},
\end{equation}
and
\begin{equation}
\left|\frac{dw}{dz}\right|_{z_1}=\frac{1}{\sqrt{\epsilon^2-\tau^2}}.
\end{equation}
With an analytical continuation $\tau\to it$, one can obtain
\begin{equation}
\left\langle \mathcal{T}^2_n(z_1)\right\rangle=\tilde{c}_n
\left(\frac{a\epsilon}{2(\epsilon^2+t^2)}\right)^{\Delta^{(2)}_{n}}.
\end{equation}
As discussed in Refs.\ \cite{Calabrese2012a,Calabrese2012b}, the scaling dimension $\Delta^{(2)}_{n}$ depends on the parity of $n$ as
\begin{equation}
\Delta_n^{(2)}=\left\{
\begin{split}
&\Delta_n\ \ \ \ &\text{odd}\  n,\\
&2\Delta_{n/2}  &\text{even}\  n,
\end{split}
\right.
,
\end{equation}
where
\begin{equation}\label{ScalingDim}
\Delta_n=\frac{c}{12}\left(n-\frac{1}{n}\right).
\end{equation}
Then by using the expressions in Eqs. (\ref{NegativityDef}) and (\ref{Disjoint}), one can get
\begin{equation}\label{E001}
\mathcal{E}=-\frac{c}{4}\ln\frac{a\epsilon}{2(\epsilon^2+t^2)}+\tilde{c}_1',
\end{equation}
where $\tilde{c}_1'=\ln\tilde{c}_1$.
As in Refs.\ \cite{Calabrese0905,Calabrese2007},
the short time behavior of $\mathcal{E}(t)$ allows us to fix the regulator $\epsilon$
in terms of the non-universal constant $\tilde{c}_1$ by requiring
\begin{equation}
\mathcal{E}(t=0)=-\frac{c}{4}\ln\frac{a}{2\epsilon}+\tilde{c}_1'=0,
\end{equation}
based on which one gets
\begin{equation}
\epsilon=\frac{a}{2}e^{-4\tilde{c}'_1/c}.
\end{equation}
Now it is possible to eliminate $a$ and $\tilde{c}'_1$ in Eq.\ (\ref{E001}) in terms of $\epsilon$, and then one can get
\begin{equation}
\mathcal{E}=-\frac{c}{4}\ln \frac{\epsilon^2}{\epsilon^2+t^2}.
\end{equation}
In the limit $t\gg\epsilon$, one ends with
\begin{equation}\label{SemiInfR}
\mathcal{E}=\frac{c}{2}\ln \frac{t}{\epsilon},
\end{equation}
which was observed in the numerical calculations based on a critical harmonic chain
\cite{Viktor}.

\subsubsection{Symmetric finite intervals}

In this part, we consider the case of symmetric finite intervals with $l_1=l_2=l$, \emph{i.e.}, $A_1\in[-l,0]$ and $A_2\in(0,l]$,
as shown in Fig.\ \ref{dL}.
In this case, $\rho_{A_1\cup A_2}$ represents a mixed state, and there is no correspondence between the logarithmic negativity and the Renyi entropies.
By using Eq.\ (\ref{Adjacent}) and doing a conformal mapping onto the RHP, one has
\begin{equation}\label{ThreePoint}
\begin{split}
\text{Tr}\left(\rho_{A_1\cup A_2}^{T_2}\right)^n
=&\left\langle \mathcal{T}_n(z_1)\bar{\mathcal{T}}_n^2(z_2) \mathcal{T}_n(z_3) \right\rangle\\
=&\prod_{i=1}^3\left|\frac{dw}{dz}\right|_{z_i}^{\Delta_{(i)}}
\left\langle \mathcal{T}_n(w_1)\bar{\mathcal{T}}_n^2(w_2) \mathcal{T}_n(w_3) \right\rangle_{\text{RHP}},
\end{split}
\end{equation}
where the scaling dimensions $\Delta_{(1)}=\Delta_{(3)}=\Delta_n$ and $\Delta_{(2)}=\Delta_n^{(2)}$.
The three-point correlation function on the RHP can be expressed as
\begin{equation}
\begin{split}
&\left\langle \mathcal{T}_n(w_1)\bar{\mathcal{T}}_n^2(w_2) \mathcal{T}_n(w_3) \right\rangle_{\text{RHP}}\\
=&\frac{\tilde{c}_n}{\prod_{i=1}^3|(w_i-\widetilde{w}_i)/a|^{\Delta_{(i)}}}\left(
\frac{\eta_{1,3}^{\Delta_n^{(2)}-2\Delta_n}}{\eta_{1,2}^{\Delta_n^{(2)}}\eta_{2,3}^{\Delta_n^{(2)}}}
\right)^{1/2}\mathcal{F}(\{\eta_{j,k}\}),
\end{split}
\end{equation}
where $\eta_{i,j}$ are cross ratios which can be constructed from the endpoints $w_i$ (and their images $\widetilde{w}_i$) of the intervals in the RHP
as follows
\begin{equation}\label{crossRatio}
\eta_{i,j}=\frac{(w_i-w_j)(\widetilde{w}_i-\widetilde{w}_j)}{(w_i-\widetilde{w}_j)(\widetilde{w}_i-w_j)}
\end{equation}
with $\widetilde{w}_i=-\bar{w}_i$ being the image of $w_i$ (see Fig.\ref{cM}). The nonuniversal function $\mathcal{F}(\{\eta_{j,k}\})$ depends on the full operator content of the CFT. $\mathcal{F}(\{\eta_{j,k}\})$ is usually difficult to calculate and only known for several specific CFTs and BCFTs. But it is found that in the limits $\eta_{i,j}\to0$, $1$, or $\infty$, the function $\mathcal{F}(\{\eta_{j,k}\})$ is just a constant, which follows from the long- and short-distance expansions of the correlation functions of twist operators \cite{Calabrese0905,Calabrese2007quench,Asplund,Cardy1991,Calabrese2011disjoint}. For symmetric intervals in this part, we calculate the cross ratios $\eta_{i,j}$ explicitly in the appendix. It is found that one always has $\eta_{ij}=1$ or $0$ for the cases $t\ll l$, $t=l+0^-$ and $t>l$. In other words, our results are universal for the above three cases.

By using Eqs. (\ref{NegativityDef}) and (\ref{Adjacent}), and neglecting various nonuniversal terms, we have
\begin{equation}\label{symmetricAdjacentA}
\begin{split}
\mathcal{E}=&
-\frac{c}{4}\ln \frac{\epsilon^2}{\epsilon^2+t^2}-\frac{c}{8}\ln\frac{\eta_{1,3}}{\eta_{1,2}\eta_{2,3}},\\
\end{split}
\end{equation}
which may be further expressed as
\begin{equation}\label{symmetricAdjacentB}
\mathcal{E}=
-\frac{c}{4}\ln \frac{\epsilon^2}{\epsilon^2+t^2}-\frac{c}{4}\ln\frac{w_{13}w_{1\tilde{2}}w_{2\tilde{3}}    }
{w_{1\tilde{3}}w_{12}w_{23}},
\end{equation}
where we have defined $w_{ij}=|w_i-w_j|$, $w_{i\tilde{j}}=|w_i-\widetilde{w}_j|$
and $w_{\tilde{i}\tilde{j}}=|\widetilde{w}_i-\widetilde{w}_j|$, respectively.

With the expressions of $w_{ij}$ that are calculated in the Appendix, one can obtain the entanglement negativity $\mathcal{E}$ as a function of time as follows
\begin{equation}
\mathcal{E}=
\left\{
\begin{split}
&-\frac{c}{4}\ln \frac{\epsilon^2}{\epsilon^2+t^2}-\frac{c}{4}\ln \frac{l+t}{l-t},   &t<l\\
&-\frac{c}{4}\ln \frac{\epsilon^2}{\epsilon^2+t^2}-\frac{c}{4}\ln\frac{4t^2}{\epsilon l},  &t>l.\\
\end{split}
\right.
\end{equation}
In the limit $l, t\gg \epsilon$, $\mathcal{E}$ can be simplified as
\begin{equation}\label{AdjacentSymmetricR}
\mathcal{E}=
\left\{
\begin{split}
&\frac{c}{4}\ln \frac{t^2+\epsilon^2}{\epsilon^2}\frac{(l-t)}{(l+t)},   &t<l\\
&\frac{c}{4}\ln \frac{l}{2\epsilon}+\text{const}, &t>l.\\
\end{split}
\right.
\end{equation}
Shown in Fig.\ \ref{SymmetricAdjacentFig}(a) is the plot of $\mathcal{E}$ with different $l$.
At the very beginning of the local quench $t\ll l$, based on Eq.\ (\ref{AdjacentSymmetricR}), one has
\begin{equation}
\mathcal{E}=\frac{c}{2}\ln\frac{t}{\epsilon},
\end{equation}
which agrees with the result of semi-infinite intervals as shown in Eq.\ (\ref{SemiInfR}).
This is reasonable because in the limit $t\ll l$,
the quasiparticles essentially propagate without noticing the finite size effect.
For $t< l$, the entanglement saturates for a certain time. Then for $t>l$, we get the ground-state value of $\mathcal{E}$, \emph{i.e.},
\begin{equation}
\mathcal{E}_G=\frac{c}{4}\ln\frac{l}{2\epsilon},
\end{equation}
which is also observed in the numerical calculations in Ref.\ \cite{Viktor}.
Note that in the numerical calculations, $\mathcal{E}$ tends to the ground-state value gradually.
In our CFT results, $\mathcal{E}$ drops to the ground-state value immediately after $t=l$.
This is because all quasiparticles propagate at the same velocity in CFTs.
In lattice models, however, the dispersion relation is not linear for all momentum vectors,
and therefore not all quasiparticles propagate at the same velocity, as discussed in detail in section IV.

\begin{figure*}
\includegraphics[width=6.25in]{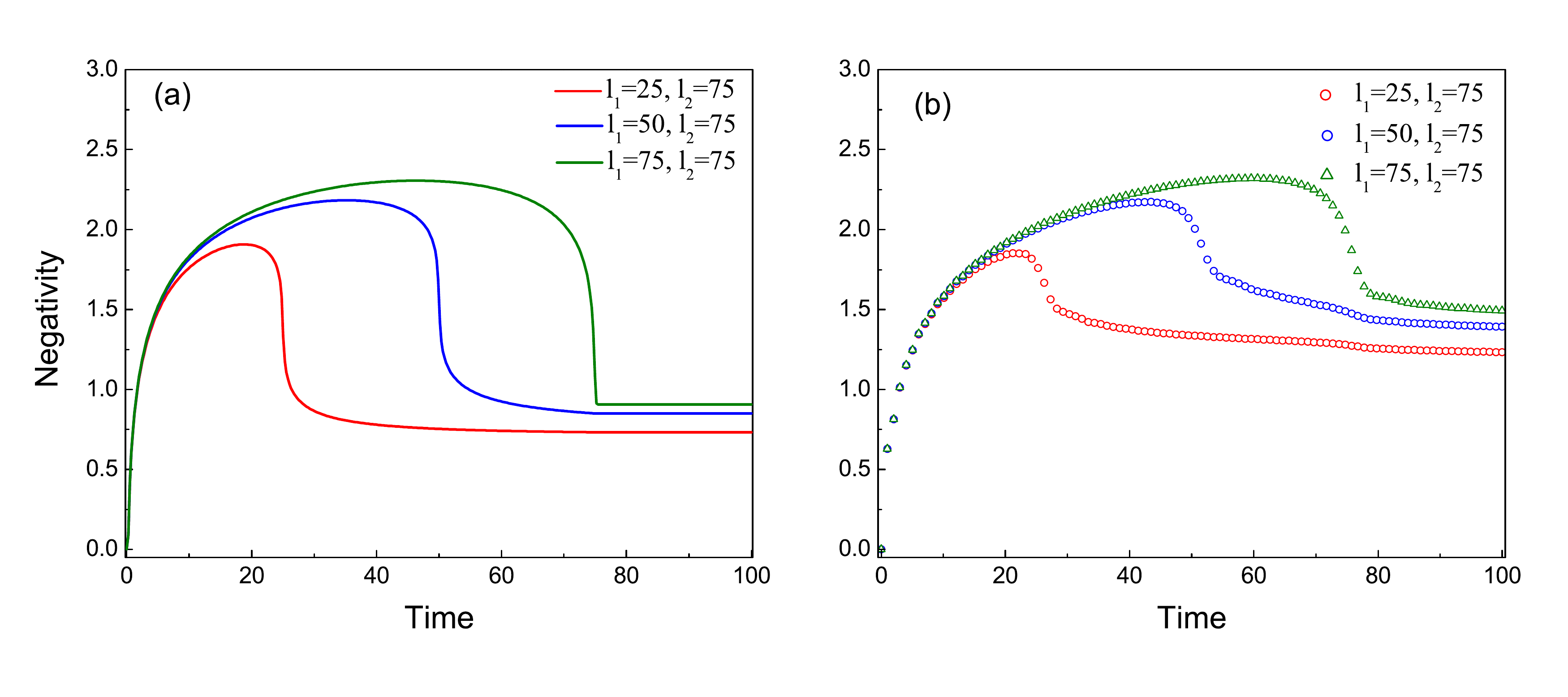}
\caption{Entanglement negativity $\mathcal{E}$ for two asymmetric adjacent intervals as a function of time. Here we choose central charge $c=1$, $\epsilon=0.1$, $(l_1,l_2)=(25,75)$, $(50,75)$, and $(75,75)$, respectively. Shown in (a) is the CFT result, and (b) is the numerical calculation based on a critical harmonic chain.}\label{AsymmetricAdjacent}
\end{figure*}

In addition,
the scaling behavior of $\mathcal{E}$ for a harmonic chain was
numerically studied in Ref.\ \cite{Viktor}.
For $t<l$, they proposed the ansatz
\begin{equation}\label{scale}
\mathcal{E}\sim \ln\frac{t^{\alpha}}{l^{-{\rho}}}\frac{(l-t)^{\beta}}{(l+t)^{-{\gamma}}}.
\end{equation}
By fitting the numerical results, they found $\alpha=1/2$, $\beta\simeq 0.15$, $\gamma\simeq 0.13$ and $\rho=-(\beta+\gamma)\simeq-0.28$. For our CFT results in Eq.\ (\ref{AdjacentSymmetricR}), by setting $c=1$ and taking the limit $t\gg \epsilon$, one has $\alpha=1/2$, $\beta= 0.25$, $\gamma= -0.25$ and $\rho=-(\beta+\gamma)= 0$. On the other hand, in the limit $t\ll l$, both Eq.\ (\ref{AdjacentSymmetricR}) and Eq.\ (\ref{scale}) collapse to $\mathcal{E}=\frac{1}{2}\ln t$.
We attribute the above disagreement/agreement to
the following fact.
For $t\le l$, because we neglect the non-universal functions $\mathcal{F}(\{\eta_{j,k}\})$  which may not be constants, our results are not accurate and therefore may not obtain the correct scaling behavior.
For $t\ll l$, however, our CFT results are universal and independent of the specific CFT.
To reproduce the numerical results for $t\le l$ in Ref.\ \cite{Viktor},
we have to consider the nonuniversal functions $\mathcal{F}(\{\eta_{j,k}\})$,
which is a difficult task, and out of the scope of our work.

Nevertheless, by comparing the values of the plateau between CFT results and numerical results in Fig.\ \ref{SymmetricAdjacentFig}, it is found they are very close to each other. To be concrete, let's take $\mathcal{E}(t=\frac{l}{2})$ for example. Based on Eq.\ (\ref{AdjacentSymmetricR}), one can get
\begin{equation}
\mathcal{E}_{\text{CFT}}(t=\frac{l}{2})=\frac{1}{2}\ln \frac{l}{2\epsilon}+\text{const}.
\end{equation}
On the other hand, from the scaling behavior in Ref.\ \cite{Viktor}, one can find
\begin{equation}
\mathcal{E}_{\text{numerical}}(t=\frac{l}{2})=\frac{1}{2}\ln \frac{l}{2}+0.13\ln 3-0.28\ln 2.
\end{equation}
In the large $l$ limit, \emph{i.e.}, $l\gg \epsilon$, one always has
\begin{equation}
\mathcal{E}_{\text{CFT}}(t=\frac{l}{2})\simeq\mathcal{E}_{\text{numerical}}(t=\frac{l}{2})\simeq\frac{1}{2}\ln \frac{l}{2}.
\end{equation}

Before we end this part, we mention that it is interesting to check how $\mathcal{E}(t)$ behaves in other lattice models.
Considering $\mathcal{E}(t<l)$ depends on the non-universal functions $\mathcal{F}(\{\eta_{j,k}\})$,
which varies for different CFTs,
we expect that for other critical lattice models such as the critical Ising model one may observe different scaling behaviors in $\mathcal{E}(t<l)$.

\subsubsection{Asymmetric finite intervals}

In this part, we consider the case of asymmetric finite intervals with $A_1\in[-l_1,0]$ and $A_2\in(0,l_2]$, as shown in Fig.\ \ref{dL}.
Without loss of generality, we suppose $l_1<l_2$.
The calculations are similar to the symmetric case, and we need to evaluate the three point correlation functions in
Eq.\ (\ref{ThreePoint}). First, as shown in the appendix, we calculate the cross ratio $\eta_{ij}$ explicitly. It is found that
one always has $\eta_{ij}=1$ or $0$ for the cases $t\ll l_1$, $t=l_2+0^-$ and $t>l_2$.
That is to say, our results are universal in these regions.
Second, by neglecting various nonuniversal terms, we arrive at the same result as
in Eq.\ (\ref{symmetricAdjacentB}).
The difference is that for asymmetric intervals, we have different expressions of $w_{ij}$, as explicitly
given in the appendix.
By plugging $w_{ij}$ into Eq.\ (\ref{symmetricAdjacentB}), one obtains
\begin{widetext}
\begin{equation}
\mathcal{E}=
\left\{
\begin{split}
&-\frac{c}{4}\ln \frac{\epsilon^2}{\epsilon^2+t^2}-\frac{c}{8}\ln \frac{(l_1+t)(l_2+t)}{(l_1-t)(l_2-t)},   &t<l_1\\
&-\frac{c}{4}\ln \frac{\epsilon^2}{\epsilon^2+t^2}-\frac{c}{4}\ln\left(\frac{2}{\epsilon}\sqrt{\frac{(l_1+l_2)(t+l_1)(t-l_1)t^2}{(l_2-l_1)l_1^2}}\right), &l_1<t<l_2\\
&-\frac{c}{4}\ln \frac{\epsilon^2}{\epsilon^2+t^2}-\frac{c}{4}\ln \frac{2(l_1+l_2)t^2}{\epsilon l_1l_2},   &t>l_2.\\
\end{split}
\right.\nonumber
\end{equation}
\end{widetext}
In the limit $l,t\gg \epsilon$, $\mathcal{E}$ can be simplified as
\begin{equation}\label{AsymmetricAdjacentR1}
\mathcal{E}=
\left\{
\begin{split}
&\frac{c}{8}\ln \frac{(l_1-t)(l_2-t)t^4}{(l_1+t)(l_2+t)\epsilon^4},   &t<l_1\\
&\frac{c}{8}\ln\frac{(l_2-l_1)l_1^2t^2}{(l_1+l_2)(t+l_1)(t-l_1)}, &l_1<t<l_2\\
&\frac{c}{4}\ln \frac{l_1l_2}{\epsilon(l_1+l_2)}+\text{const},   &t>l_2.\\
\end{split}
\right.
\end{equation}
Note that in the limit $l_1=l_2$, we can reproduce the symmetric adjacent intervals result in Eq.\ (\ref{AdjacentSymmetricR}).
The plot of $\mathcal{E}$ for different $(l_1,l_2)$ is shown in Fig.\ \ref{AsymmetricAdjacent}(a). First, for $t\ll \min[l_1,l_2]$,
one can find the time evolution of $\mathcal{E}(t)$ is similar to the symmetric case. Actually, based on Eq.\ (\ref{AsymmetricAdjacentR1}), one can check
that in the limit $t\ll \min[l_1,l_2]$
\begin{equation}
\mathcal{E}=\frac{c}{2}\ln \frac{t}{\epsilon},
\end{equation}
which shows the $\ln t$ behavior again, as expected. For $t>\max[l_1,l_2]$,
we obtain the ground state value of the entanglement negativity
\cite{Calabrese2012a,Calabrese2012b}, i.e.,
\begin{equation}
\mathcal{E}_G=\frac{c}{4}\ln\frac{l_1l_2}{\epsilon(l_1+l_2)}.
\end{equation}
Interestingly, it is found that the sudden drop of $\mathcal{E}$ happens at
\begin{equation}
t=\min[l_1,l_2],
\end{equation}
which is again straightforward to understand based on
the heuristic physical picture that the quasiparticles may be viewed as entanglement pairs of two quanta.
For $t<\min[l_1,l_2]$, two entangled quasiparticles are in $A_1$ and $A_2$ separately, and create the entanglement between $A_1$ and $A_2$. At $t=\min[l_1,l_2]=l_1$ (here we suppose $l_1<l_2$), although one quasiparticle is still in $A_2$, the other quasiparticle propagates out of $A_1$, and therefore the entanglement between $A_1$ and $A_2$ decreases suddenly at $t=\min[l_1,l_2]$.

Before we end this part, we emphasize that our results are universal for the regimes
$t\ll l_1$, $t=l_2+0^-$ and $t>l_2$.
For the case of $t\le l_2$, however,
similar with the symmetric case,
because we neglected the nonuniversal functions $\mathcal{F}(\{\eta_{j,k}\})$
which may not be constants,
our results are not accurate, and one has to calculate $\mathcal{F}(\{\eta_{j,k}\})$ for different CFTs.

\begin{figure*}
\includegraphics[width=6.25in]{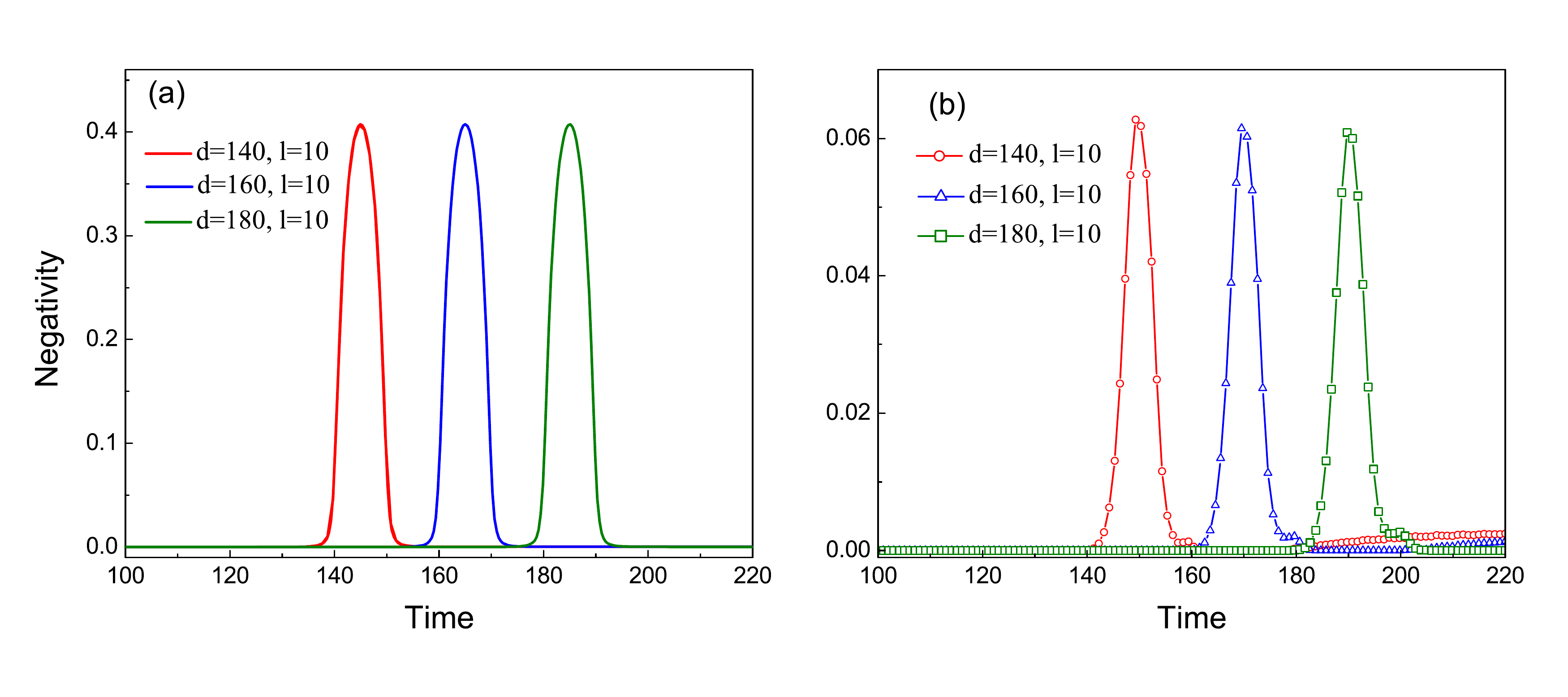}
\caption{Entanglement negativity $\mathcal{E}$ for two symmetric disjoint intervals as a function of time.
Here we choose the central charge $c=1$, $\epsilon=1$, $(d,l)=(140, 10)$, $(160,10)$ and $(180, 10)$, respectively. Shown in (a) is the CFT result, and (b) is the numerical calculation based on a critical harmonic chain. }\label{SymmetricDisjoint}
\end{figure*}

\subsection{Two disjoint intervals }

\subsubsection{Symmetric finite intervals}

In this part, we consider the symmetric disjoint intervals, \emph{i.e.}, $A_1\in[-d-l,-d]$ and $A_2\in[d,d+l]$, as shown Fig.\ \ref{dL}(b).
In this case, we need to consider the correlation function of four twist fields as shown in Eq.\ (\ref{Disjoint}). By applying the conformal map in Eq.\ (\ref{conformalMap}), one has
\begin{equation}
\begin{split}
&\left\langle \mathcal{T}_n(z_1)\bar{\mathcal{T}}_n(z_2)\bar{\mathcal{T}}_n(z_3) \mathcal{T}_n(z_4) \right\rangle\\
=&\prod_{i=1}^4\left|\frac{dw}{dz}\right|_{z_i}^{\Delta_n}
\left\langle \mathcal{T}_n(w_1)\bar{\mathcal{T}}_n(w_2)\bar{\mathcal{T}}_n(w_3) \mathcal{T}_n(w_4) \right\rangle_{\text{RHP}},
\end{split}\nonumber
\end{equation}
where the four-point correlation function on the RHP has the form
\begin{equation}
\begin{split}
&\left\langle \mathcal{T}_n(w_1)\bar{\mathcal{T}}_n(w_2)\bar{\mathcal{T}}_n(w_3) \mathcal{T}_n(w_4) \right\rangle_{\text{RHP}}\\
=&\frac{\tilde{c}_n^2}{\prod_{i=1}^4|(w_i-\widetilde{w}_i)/a|^{\Delta_n}}
\frac{1}{\eta_{1,2}^{\Delta_n}\eta_{3,4}^{\Delta_n}}
\left(\frac{\eta_{1,4}\eta_{2,3}}{\eta_{1,3}\eta_{2,4}}\right)^{\Delta_n^{(2)}/2-\Delta_n}\\
&\times \mathcal{F}\left(\{\eta_{j,k}\}\right).
\end{split}
\end{equation}
For the nonuniversal functions $\mathcal{F}(\{\eta_{j,k}\})$, as explicitly calculated in Ref.\ \cite{Asplund}, they are simply a constant in the limit $l/d\ll 1$. In other words, when the two intervals are far apart, we do not need the knowledge of $\mathcal{F}\left(\{\eta_{j,k}\}\right)$.
By using the definition in Eq.\ (\ref{NegativityDef}), and dropping various
multiplicative constants, we have
\begin{equation}\label{ENdis}
\mathcal{E}=-\frac{c}{8}\ln\left(\frac{\eta_{1,4}\eta_{2,3}}{\eta_{1,3}\eta_{2,4}}\right),
\end{equation}
which is alternatively written as
\begin{equation}\label{Negativity001}
\begin{split}
\mathcal{E}
=&-\frac{c}{8}\ln\frac{w_{14}w_{1\tilde{3}}w_{23}w_{2\tilde{4}} w_{\tilde{1}\tilde{4}}w_{\tilde{1}3}w_{\tilde{2}\tilde{3}}w_{\tilde{2}4}    }{w_{1\tilde{4}}w_{13}w_{2\tilde{3}}w_{24}w_{\tilde{1}4} w_{\tilde{1}\tilde{3}}w_{\tilde{2}3}w_{\tilde{2}\tilde{4}} }.
\end{split}
\end{equation}
By noting that $w_{ij}=w_{\tilde{i}\tilde{j}}$ and $w_{i\tilde{j}}=w_{\tilde{i}j}$, Eq.\ (\ref{Negativity001}) can be simplified as
\begin{equation}\label{Negativity002}
\begin{split}
\mathcal{E}
=&-\frac{c}{4}\ln\frac{w_{14}w_{1\tilde{3}}w_{23}w_{2\tilde{4}} }{w_{1\tilde{4}}w_{13}w_{2\tilde{3}}w_{24}}.
\end{split}
\end{equation}
With the explicit forms of $w_{ij}$ given in the Appendix, we can obtain the entanglement negativity $\mathcal{E}$ as a function of time as follows
\begin{equation}\label{SymmetricDisR}
\mathcal{E}=
\left\{
\begin{split}
&0   &t<d\\
&\frac{c}{4}\ln\frac{(2d+l)(d+l-t)(t^2-d^2)}{\epsilon dl(d+l+t)}   &d<t<d+l\\
&\frac{c}{4}\ln\frac{(2d+l)^2}{4d(d+l)}  &t>d+l
\end{split}
\right.
\end{equation}
Note that in the study of the negativity evolution after a global quench, it was found that $\mathcal{E}(t)$ shows the
same behavior as the Renyi mutual information apart from the prefactor
\cite{Coser}.
For the local quench studied here, by comparing our result in Eq.\ (\ref{SymmetricDisR}) with the result of mutual information in Ref.\ \cite{Asplund},
it is found that the expressions are also the same except for the prefactor.
In other words, our results parallel with the story in the negativity evolution after a global quench.
The relation between the entanglement negativity and the mutual information after a local quantum quench will be systematically discussed in section IV.

\begin{figure*}
\includegraphics[width=6.25in]{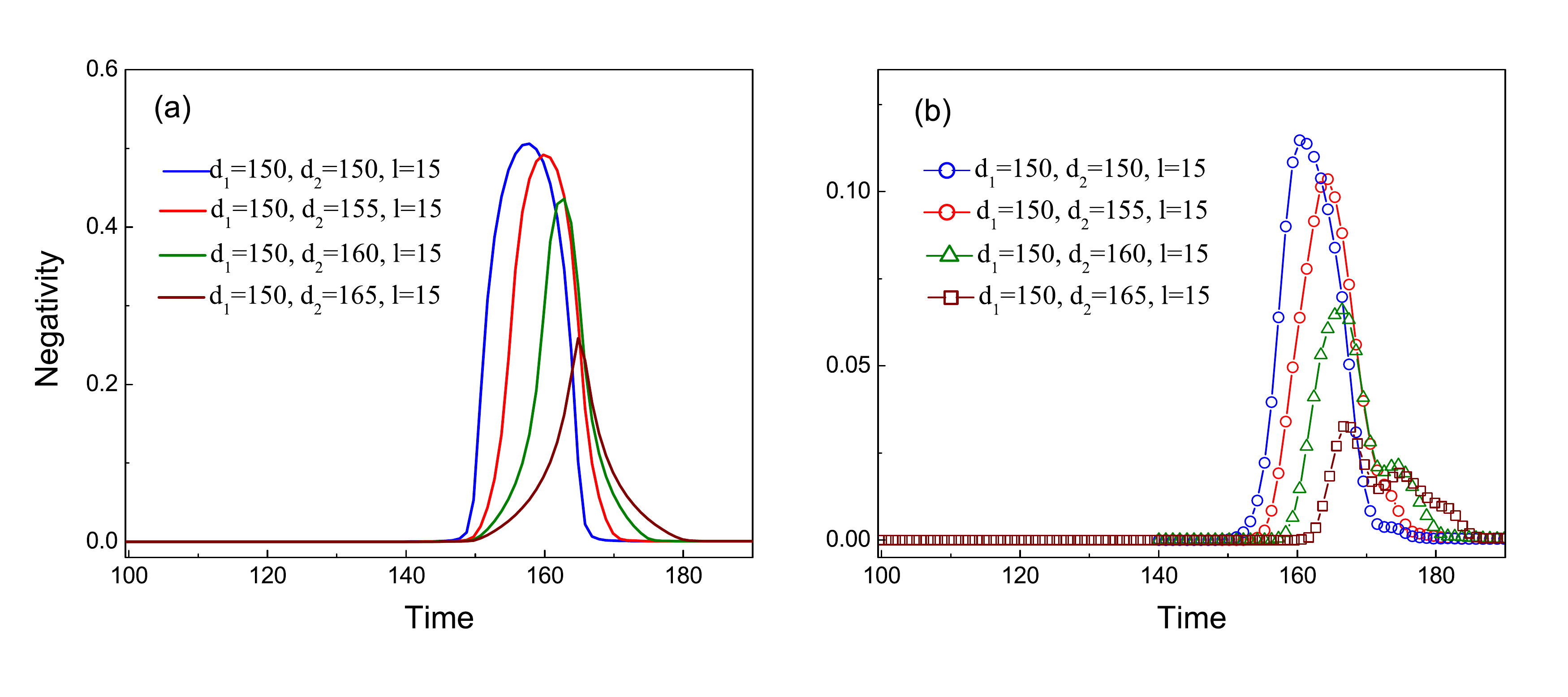}
\caption{Entanglement negativity $\mathcal{E}$ for two asymmetric disjoint intervals as a function of time.
Here we choose the central charge $c=1$, $\epsilon=1$, $l=15$, $(d_1,d_2)=(150, 150)$, $(150, 155)$, $(150,160)$, and $(150,165)$, respectively. Shown in (a) is the CFT result, and (b) is the numerical calculation based on a critical harmonic chain. }\label{AsymmetricDisjoint}
\end{figure*}

As shown in Fig.\ \ref{SymmetricDisjoint}(a), we plot the evolution of the entanglement negativity with different $(d,l)$ according to Eq.\ (\ref{SymmetricDisR}).
A `light-cone' effect can be observed: For $t<d$, there is no entanglement negativity between $A_1$ and $A_2$. At $t=d$, the entanglement negativity begins to develop,
and reaches the maximum approximately at $t=d+l/2$. At $t=d+l$, the entanglement negativity decreases suddenly, which corresponds to the entangled pairs propagating out of intervals
$A_1$ and $A_2$ simultaneously. Note that at $t=d+l/2$, taking the limit $d\gg l$, one has
\begin{equation}
\mathcal{E}_{t=d+\frac{l}{2}}\simeq\frac{c}{4}\ln\frac{l}{2\epsilon},
\end{equation}
which is \emph{independent} of the distance $d$, as also can be observed in Fig.\ \ref{SymmetricDisjoint}.
That is to say, with the help of entangled pairs, we can create a long-range entanglement between two intervals which are far from each other.
Note that this long-range entanglement was also observed in the time evolution of mutual information $I(t)$ in Ref.\ \cite{Asplund}, where it is found that $I_{t=d+\frac{l}{2}}\simeq\frac{c}{3}\ln\frac{l}{2\epsilon}$.

%
%
\subsubsection{Asymmetric finite intervals}

In this part, we consider the asymmetric disjoint intervals.
We have multiple choices as follows: (i) $d_1\neq d_2$, $l_1=l_2$, (ii) $d_1= d_2$, $l_1\neq l_2$
and (iii) $d_1\neq d_2$, $l_1\neq l_2$.
For simplicity, we consider the case (i), \emph{i.e.}, $A_1\in[-d_1+l,-d_1]$ and $A_2\in[d_2,d_2+l]$. Without loss of generality, we choose $d_1<d_2\le d_1+l$.

The calculation of the negativity evolution is similar with the symmetric case,
and we obtain the same result in Eq.\ (\ref{Negativity002}).
The difference is that we should express $w_{ij}$ in terms of $d_1$, $d_2$ and $l$,
as explicitly shown in the appendix.
By plugging the expressions of $w_{ij}$ into Eq.\ (\ref{Negativity002}),
one arrives at the time evolution of the entanglement negativity
\begin{widetext}
\begin{equation}\label{unsymmetricDisjoint}
\mathcal{E}=
\left\{
\begin{split}
&0  &t<d_1\\
&-\frac{c}{8}\ln\frac{(d_1+d_2)(d_2-t)(d_2+l+t)(d_2-d_1+l)}{(d_1+d_2+l)(d_2+l-t)(d_2+t)(d_2-d_1)}    &d_1<t<d_2\\
&-\frac{c}{4}\ln\frac{\epsilon(d_1+d_2)}{2(d_1+d_2+l)}\sqrt{\frac{(d_1+l-d_2)(d_2+l-d_1)(d_1+l+t)(d_2+l+t)}{(d_1+l-t)(d_2+l-t)(t^2-d_1^2)(t^2-d_2^2)}}  &d_2<t<d_1+l\\
&-\frac{c}{8}\ln\frac{[t^2-(d_1+l)^2](d_1+d_2)^2(d_2+l-d_1)(d_1+d_2+2l)}{(d_2-d_1)(d_1+d_2+l)^3(t^2-d_1^2)}  &d_1+l<t<d_2+l\\
&-\frac{c}{4}\ln\frac{(d_1+d_2+2l)(d_1+d_2)}{(d_1+d_2+l)^2}   &t>d_2+l\\
\end{split}
\right.
\end{equation}
\end{widetext}
One can check that when $d_1=d_2=d$, the result in Eq.\ (\ref{SymmetricDisR}) is reproduced.

According to Eq.\ (\ref{unsymmetricDisjoint}), we plot $\mathcal{E}(t)$ with different $(d_1,d_2)$ in Fig.\ \ref{AsymmetricDisjoint}(a).
Compared to the symmetric case, the `light-cone' effect is still observed. The difference is that the time when $\mathcal{E}(t)$ increases quickly now happens at \begin{equation}
t=\max[d_1,d_2],
\end{equation}
and the time when $\mathcal{E}(t)$ decreases quickly happens at \begin{equation}
t=\min[d_1+l,d_2+l],
\end{equation}
which is also in agreement with the quasiparticle picture.

\section{Numerical evaluation of the negativity for a harmonic chain after a local quench}
In this section, to confirm our CFT results,
we study the time evolution of the logarithmic negativity after a local quantum quench
in a lattice model,
a critical harmonic chain.
The entanglement negativity for a harmonic chain has been numerically studied in several works
\cite{Audenaert2002, Botero, Calabrese2012b, Viktor, Coser}.
Here we follow the method developed in these works, and apply it to the local quench problem.
We will first introduce the lattice model and the covariance matrix in part A. In part B,
we introduce the evolution matrix and show how to calculate the entanglement negativity.
In part C, we apply the method to the cases studied with CFT approach, and compare the results accordingly.

\subsection{Harmonic chain and the covariance matrix}
The Hamiltonian of the harmonic chain is
\begin{align}
H=\sum_{n=1}^{N}[ \frac{p^2_n}{2 M} + \frac{M \omega_0^2}{2}q_n^2 + \frac{K}{2} (q_{n+1}-q_n)^2],
\end{align}
where $N$ is the number of sites of the chain, $M$ is the mass scale, $\omega_0$
is the characteristic frequency, and $K$ is the nearest-neighbor coupling.
$p_n$ and $q_n$ denote the momentum and position operators with canonical commutation
relations $[p_n,p_m]=[q_n,q_m]=0$ and $[q_n, p_m]=i \delta_{n,m}$.

For periodic boundary condition (PBC),  the Fourier transform
of the canonical variables are
\begin{equation}
\left\{
\begin{split}
q_n=&\sum_{k=1}^{L} \widetilde{q}_k \frac{1}{\sqrt{L}}e^{2 \pi i k n /L},\\
\quad \widetilde{q}_k=&\sum_{n=1}^{L} q_{n} \frac{1}{\sqrt{L}}e^{-2 \pi i k n /L},
\end{split}
\right.
\end{equation}
where $n=1,\cdots, L$. For $p_n$, the Fourier transform is identical to $q_n$.
The Hamiltonian is diagonalized in the momentum space
\begin{align}
H=\sum_{k=1}^{L}\left ( \frac{1}{2M}  \widetilde{p}_k^2+\frac{M \omega_k^2}{2} \widetilde{q}_k^2 \right ),
\label{Eq: HC}
\end{align}
where
\begin{align}\label{dispersion1}
\omega_k=\sqrt{\omega_0^2 + \frac{4 K}{M} \sin (\frac{\pi k}{L})^2}, \quad k=1,\cdots, L, \quad (\mathrm{PBC}).
\end{align}

For the Dirichlet boundary condition (DBC), the Fourier transform is not valid due to the breaking of translational symmetry.
However, the Fourier sine transform can be defined as
\begin{equation}
\left\{
\begin{split}
q_n=&\sum_{k=1}^{L-1}  \widetilde{q}_k  \sqrt{\frac{2}{L}} \sin \left (   \frac{\pi k n}{L} \right ), \\
 \widetilde{q}_k=& \sum_{n=1}^{L-1} q_{n} \sqrt{\frac{2}{L}} \sin \left( \frac{\pi k n}{L} \right ),
\end{split}
\right.
\end{equation}
where $n=1,\cdots, L$.
For $p_n$, the Fourier sine transformation is defined similarly.
The Hamiltonian in the momentum space is identical to Eq. (\ref{Eq: HC}).
But the frequency $\omega_k$ has a different form,
\begin{align}\label{dispersion2}
\omega_k=\sqrt{\omega_0^2 + \frac{4 K}{M} \sin (\frac{\pi k}{2L})^2}, \quad k=1,\cdots, L-1, \quad (\mathrm{DBC}).
\end{align}

The covariance matrix is constructed from the  two-point correlators
\begin{align}
\gamma_{n,m} = \mathrm{Re} \left(\begin{array}{cc}\langle 0|q_n q_m|0 \rangle & \langle 0|q_n p_m|0 \rangle \\\langle 0|p_n q_m|0 \rangle & \langle 0|p_n p_m|0 \rangle\end{array}\right).
\label{Eq: CM}
\end{align}
For PBC, the correlators are
\begin{align}
\langle 0|q_n q_m|0 \rangle &=\frac{1}{2L} \sum_{k=1}^{L} \frac{1}{M \omega_k} \cos \left [ \frac{2 \pi k (n-m)}{L} \right ], \notag\\
\langle 0|p_n p_m|0 \rangle &=\frac{1}{2L} \sum_{k=1}^{L} M \omega_k \cos \left [ \frac{2 \pi k (n-m)}{L} \right ], \notag\\
\langle 0|q_n p_m|0 \rangle &=i \delta_{n,m}/2.
\end{align}
For DBC, the correlators are
\begin{align}
\langle 0|q_n q_m|0 \rangle &=\frac{1}{L} \sum_{k=1}^{L-1} \frac{1}{M \omega_k} \sin \left (\frac{\pi k n}{L} \right ) \sin \left (\frac{\pi k m}{L} \right ), \notag\\
\langle 0|p_n p_m|0 \rangle &=\frac{1}{L} \sum_{k=1}^{L-1} M \omega_k \sin \left (\frac{\pi k n}{L} \right ) \sin \left (\frac{\pi k m}{L} \right ), \notag\\
\langle 0|q_n p_m|0 \rangle &=i \delta_{n,m}/2.
\label{Eq:DBC}
\end{align}

\subsection{Evolution matrix and the logarithmic negativity}
From the Heisenberg equation of motion, $\dot{ \widetilde{q}}_k(t)=\frac{1}{M} \widetilde{p}_k(t)$ and
$\dot{ \widetilde{p}}_k(t)=-M \omega_k^2 \widetilde{q}_k(t)$, we have
\begin{equation}
\left\{
\begin{split}
 \widetilde{q}_k(t)&=\frac{1}{\sqrt{M}} (\cos \omega_k t  \widetilde{q}_k(0) + \omega_k^{-1} \sin \omega_k t \widetilde{p}_k(0)), \notag\\
 \widetilde{p}_k(t)&=\sqrt{M}(-\omega_k \sin \omega_k t  \widetilde{q}_k(0) + \cos \omega_k t  \widetilde{p}_k(0)).
\end{split}
\right.
\end{equation}
The time-dependent canonical variables in the real space are
\begin{equation}
\left\{
\begin{split}
q_n(t)=&\frac{1}{\sqrt{M}}\sum_{k,m} \phi^*_k(n) \phi_k(m)\\
&\times\left (\cos \omega_k t q_m(0) + \omega_k^{-1} \sin \omega_k t p_m(0) \right ) \notag\\
p_n(t)=&\sqrt{M}\sum_{k,m}\phi^*_k(n) \phi_k(m) \\
&\times\left (-\omega_k \sin \omega_k t q_m(0) + \cos \omega_k t p_m(0) \right ),
\end{split}
\right.
\end{equation}
where
\begin{align}
\phi_n(k)&= \frac{1}{\sqrt{L}}e^{-2 \pi i k n /L}, \quad  (\mathrm{PBC}), \notag\\
 \phi_n(k)&= \frac{2}{\sqrt{L}}\sin  (\pi  k n /L),  \quad (\mathrm{DBC}).
\end{align}
Therefore, the time evolution of the covariance matrix is
\begin{equation}\label{Eq: CMt}
\gamma(t) = S (t) \gamma (0) S(t)^\mathrm{T},
\end{equation}
where the evolution matrix is
\begin{equation}\label{Eq: SM}
\begin{split}
S_{n,m} (t)=&\sum_{k} \phi^*_k(n) \phi_k(m)\\
&\times\left(\begin{array}{cc}  \frac{1}{\sqrt{M}} \cos \omega_k t  & \frac{1}{\sqrt{M}} \omega_k^{-1} \sin \omega_k t  \\
 -\sqrt{M}\omega_k \sin \omega_k t  & \sqrt{M}\cos \omega_k t \end{array}\right).
\end{split}
\end{equation}

The entanglement properties are encoded in the reduced density matrix,
which can be extracted from the the covariance matrix $\gamma_A$ associated with the the subsystem $A$.
The logarithmic negativity is defined by the partial transposition of the reduced density matrix $\rho_A$ with
the subsystem $A=A_1 \cup A_2$ as $\mathcal{E} = \mathrm{ln} \mathrm{Tr}|\rho_A^{T_2}|$.
We first consider the partial transposition of $\gamma_A$, which can be constructed by inverting
the signs of the momenta corresponding to $A_2$ \cite{Audenaert2002}.
\begin{align}
\gamma_A^{T_2}  =  \left(\begin{array}{cc}\mathbb{I}_{l_A} & 0_{l_A} \\0_{l_A} & \mathbb{R}_{A_2}\end{array}\right) \cdot
\gamma_A \cdot
\left(\begin{array}{cc}\mathbb{I}_{l_A} & 0_{l_A} \\0_{l_A} & \mathbb{R}_{A_2}\end{array}\right),
\end{align}
where
\begin{align}
[\mathbb{R}_{A_2}]_{n,m} = \left(\begin{array}{cc} 1 & 0 \\0 & -1 \end{array}\right) \delta_{n,m}, \quad n,m \in A_2,
\end{align}
and $l_A$ is the number of sites in region $A$. The symplectic spectrum of $\gamma_A^{T_2}$ can be obtained by exact diagonalization after multiplying with a symplectic matrix $\Sigma$
\begin{align}
\Sigma =  \left(\begin{array}{cc} 0 & \mathbb{I}_{l_A} \\ -\mathbb{I}_{l_A} & 0 \end{array}\right).
\end{align}

After computing the spectrum of $i \Sigma \cdot \gamma_A$ with the set of eigenvalues $\{ \pm \lambda_1, \pm \lambda_2, \cdots \pm \lambda_{l_A}; \lambda_\mu >0 \}$,
the trace norm of the partial transposition of the reduced density matrix is \cite{Audenaert2002,Calabrese2012b}
\begin{align}
\mathrm{Tr}|\rho_A^{T_2}|=\prod_{\nu=1}^{l_A} \frac{1}{|\lambda_\nu+\frac{1}{2}|-|\lambda_\nu-\frac{1}{2}|}=\prod_{\nu=1}^{l_A} \mathrm{max} \left ( 1, \frac{1}{2\lambda_\mu}\right).
\end{align}
Therefore, the logarithmic negativity is
\begin{align}
\mathcal{E}= -\sum_{\nu=1}^{l_A} \mathrm{ln} \left [  \mathrm{min} ( 1, 2\lambda_\mu) \right].
\label{Eq:EN}
\end{align}

Let us summarize the details of computing the negativity:
\begin{itemize}
\item
Constructing the covariance matrix based on Eq. (\ref{Eq: CM}).
\item
Constructing the evolution matrix $S_{n,m}(t)$ [Eq. (\ref{Eq: SM})] and obtain the time-evolution covariance matrix $\gamma(t)$ [Eq. (\ref{Eq: CMt})].
\item
Constructing the reduced covariance matrix $\gamma_A(t)$ and computing the spectrum of $i \Sigma \cdot \gamma_A(t)$.
Then, the negativity is obtained by Eq. (\ref{Eq:EN}).
\end{itemize}

\subsection{Evolution of entanglement negativity after a local quench}
We consider two disconnected harmonic chains with equal number of sites $N$ under the Dirichlet boundary condition, with each harmonic chain prepared in its ground state. At $t=0$, two disconnected chains are joined together as one harmonic chain with number of sites $2N$ under the Dirichlet boundary condition
(similar to Ref. \cite{Viktor}).

From Eqs. (\ref{Eq: CM}) and (\ref{Eq:DBC}), the covariance matrix at $t=0$ is
\begin{align}
\gamma(0) = \left(\begin{array}{cc} \gamma_1(0) &0 \\  0 & \gamma_2(0)  \end{array}\right),
\end{align}
where
\begin{equation}
\begin{split}
&[\gamma_1(0)]_{n,m}=[\gamma_2(0)]_{n,m}\\
=&\frac{1}{N}\sum_{k=1}^{N-1} \sin \left( \frac{\pi k n}{N}\right) \sin \left( \frac{\pi k m}{N}\right)
 \left(\begin{array}{cc} \frac{1}{M \omega_k} &0 \\  0 & M \omega_k \end{array}\right),
\end{split}
\end{equation}
with $\omega_k=\sqrt{\omega_0^2 + \frac{4 K}{M} \sin (\frac{\pi k}{2N})^2}$,
and $\gamma_{1/2}(t)$ denotes the covariance matrix of the disconnected chain.
The evolution matrix in the situation is
\begin{equation}
\begin{split}
S_{n,m}(t)
=&\frac{1}{N}\sum_{k=1}^{2N-1} \sin \left( \frac{\pi k n}{2N}\right) \sin \left( \frac{\pi k m}{2N}\right)\\
&\times
\left(\begin{array}{cc}  \frac{1}{\sqrt{M}} \cos \Omega_k t  & \frac{1}{\sqrt{M}} \Omega_k^{-1} \sin \Omega_k t  \\
 -\sqrt{M}\Omega_k \sin \Omega_k t  & \sqrt{M}\cos \Omega_k t \end{array}\right),
\end{split}
\end{equation}
where $\Omega_k=\sqrt{\omega_0^2 + \frac{4 K}{M} \sin (\frac{\pi k}{4N})^2}$.

In the following,
the negativity is computed by setting $M=K=1$ and $\omega_0=0$
so that the maximal group velocity of normal-mode excitations
is set to unity.
(Hereafter we will use the `light-speed' $v_c (=1)$ to represent the maximal group velocity.)
The total length of the harmonic chain is $N=500$.
The partial transposition is performed with respect to the interval $A_2$.
Notice that for $\omega_0=0$, the system is critical with the central charge $c=1$.
As shown in Figs.\ \ref{SymmetricAdjacentFig}$\sim$\ref{AsymmetricDisjoint},
we compute the negativity evolution for both adjacent and
disjoint intervals, including symmetric and asymmetric cases. By comparing the analytical results obtained from CFT approach and the numerical
results based on the harmonic chain, it is the found that the main features agree very well.

\subsection{Comparison between CFT results and numerical results}

Although the main features of our CFT results and numerical results agree with each other,
we notice that there are some disagreements in detail between the two methods as follows.

\begin{figure}[htp]
\includegraphics[scale=0.47]{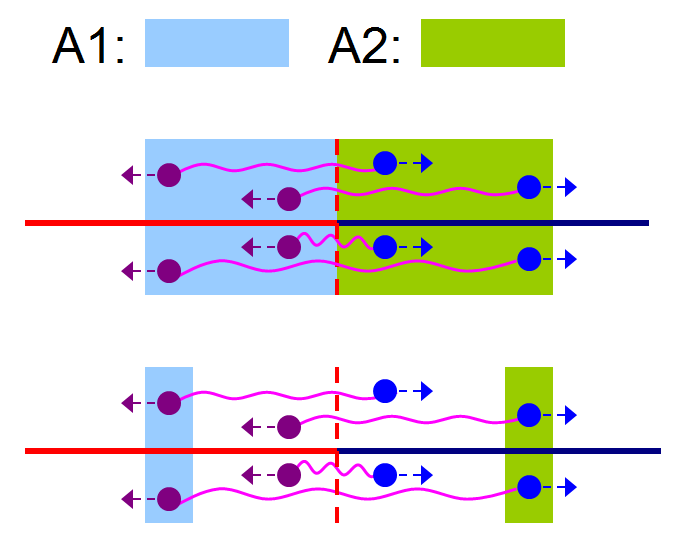}
\caption{Effects of propagating entangled pairs on the entanglement negativity $\mathcal{E}$ of two adjacent intervals and two disjoint intervals in a lattice model.   For two adjacent intervals, all the four kinds of entangled pairs may contribute to
the entanglement negativity. For two disjoint intervals, however, only the fast-fast entangled pair may contribute
to the entanglement negativity near $t\sim d$, as discussed in the main text.}\label{EP}
\end{figure}

(i) For the results of two adjacent intervals in Fig.\ \ref{SymmetricAdjacentFig},
at time $t=l/v_c=l$, the entanglement negativity obtained from
the CFT approach drops to the ground-state value suddenly.
For the lattice model, however, the entanglement negativity approaches
the ground-state value gradually.
This is due to the fact that  in CFTs all the quasiparticles propagate at the same speed $v_c$,
which is not the case in a lattice model.
As shown in Eq. (\ref{dispersion2}),
the dispersion relation is nonlinear, which indicates that not all the quasiparticles have the same group velocity.
In particular, for quasi-particles with higher energy, their group velocities $v$ are smaller than the light speed $v_c$,
\emph{i.e.}, $v<v_c$.
As shown in Fig.\ \ref{EP}, qualitatively, we can divide the entangled pairs into four groups according
to their group velocities $(v_L,v_R)$ :
\begin{equation}
(v_L,v_R)\simeq\left\{
\begin{split}
&(-v_c,v_c), \ \ \ &\text{fast-fast pair}\\
&(-v_c,v), \ \ \ &\text{fast-slow pair}\\
&(-v,v_c), \ \ \ &\text{slow-fast pair}\\
&(-v,v), \ \ \ &\text{slow-slow pair}.\\
\end{split}\right.
\end{equation}
In the CFT study of two adjacent intervals,
there are only fast-fast pairs which lead to the abrupt drop of $\mathcal{E}$ at $t=l$.
In the lattice model, however, the slow-slow pairs still make contributions to $\mathcal{E}$ even for $t>l$,
and this is why $\mathcal{E}$ drops to the ground-state value gradually in our numerical results.

(ii) The concrete values of the entanglement negativity obtained from the
CFT approach and the numerical method
do not agree with each other in a perfect way.
In particular, the CFT results are much larger than the numerical results for the disjoint cases,
as shown in Fig.\ \ref{SymmetricDisjoint} and Fig.\ \ref{AsymmetricDisjoint}.
This phenomenon, again, may be understood based on the quasi-particle picture in Fig.\ \ref{EP}.
For two symmetric adjacent intervals,
all the four kinds of entangled pairs contribute to $\mathcal{E}$ for $t<l$.
For two symmetric disjoint intervals, however, only the fast-fast pairs contribute to $\mathcal{E}$ during $d<t<d+l$,
and the other three kinds of entangled pairs do not make any contribution at all.
This explains why the numerical results are much smaller than the CFT results for the disjoint cases.

The above quasi-particle picture on a lattice model may lead to many interesting results.
For example, in the symmetric disjoint case, careful readers may wonder if the slow-slow pairs will make contributions
to $\mathcal{E}$ at a later time $t>d+l$. The answer is yes.
Actually, as shown in Fig.\ \ref{SymmetricDisjoint}, for the case $d=140$ and $l=10$,
one can find that
$\mathcal{E}$
shows
a ``tail'' or ``revival'' starting at around $t=180$,
which results from the contribution of slow-slow pairs.

\begin{figure}
\includegraphics[scale=0.46]{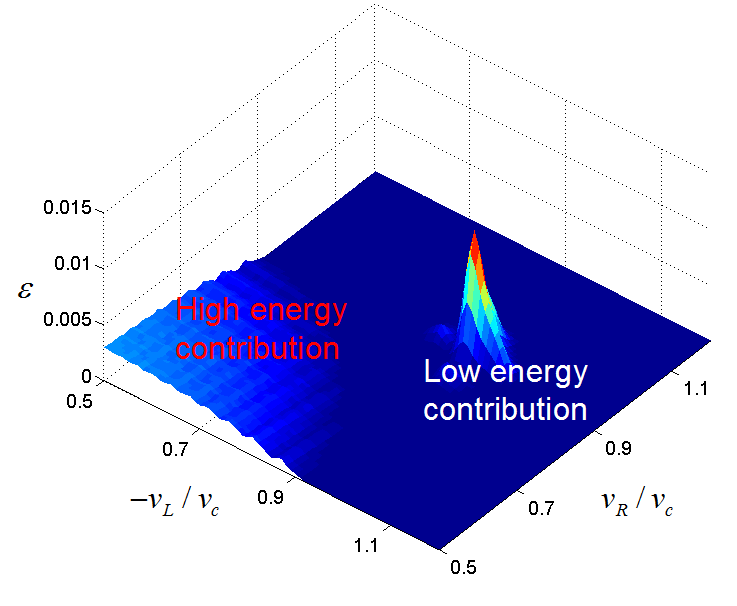}
\caption{Distribution of entangled pairs as a function of $(v_L,v_R)$ for a harmonic chain. The peak at $(-v_L/v_c, v_R/v_c)\simeq(1,1)$ is contributed by the fast-fast pairs. The region  ( $-v_L/v_c,v_R/v_c\le 1$) with finite $\mathcal{E}$
 is contributed by the other three kinds of entangled pairs, \emph{i.e.}, fast-slow pairs, slow-fast pairs and slow-slow pairs, which
result from the high energy contribution. The parameters we choose are $l_1=l_2=l=5$ and $t=100$.
}\label{EP3}
\end{figure}

To demonstrate the above physical picture,
we study, numerically,
the distribution of entangled pairs as a function of $(v_L,v_R)$ based on the harmonic chain.
The strategy is as follows.
By choosing disjoint intervals $A_1$ and $A_2$ as shown in Fig.\ref{dL}, we fix the length of two intervals $l_1=l_2=l=5$
and time $t=100$. Then we measure $\mathcal{E}$ by changing the distances $d_1$ and $d_2$ separately.
In this way, we plot $\mathcal{E}$ as a function of $(v_L/v_c,v_R/v_c)=(-d_1/t,d_2/t)$.
As shown in Fig.\ \ref{EP3}, it is found that there is a peak located at $(-v_L/v_c,v_R/v_c)\simeq (1,1)$
as expected from the CFT approach
which is mainly contributed by the fast-fast pairs.
On the other hand, there is another region ( $-v_L,v_R\le v_c$) with finite $\mathcal{E}$,
which is mainly contributed by the other three kinds of entangled pairs, \emph{i.e.}, fast-slow pairs,
slow-fast pairs and slow-slow pairs, respectively.
While suppressed as compared to the main peak at $(1,1)$,
these quasiparticle pairs still carry non-negligible contributions to the entanglement negativity,
leading to the tail or revival of the negativity $\mathcal{E}$.

Then a natural question is: how can we understand the ``shape'' of entangled pairs,
i.e., the distribution of entangled pairs as a function of $(v_L,v_R)$?
In particular, for different methods of local quenches
\cite{Calabrese2012a,Calabrese2012b,Tadashi1302, Tadashi1401,Tadashi1403},
the shape of entangled pairs may be very different.
A quantitative study of this question as well as its application to the finite temperature case will be reported elsewhere
\cite{Wen2}.

Another question that the experts may ask is: For the time evolution of the entanglement negativity for two disjoint intervals after a global quench in Ref.\cite{Coser}, why cannot we observe an apparent mismatch between the CFT results and the numerical results? The reason is that Ref.\cite{Coser} mainly focuses on the case $l\ge d$, which is close to the case of adjacent intervals. The shape of entangled pairs plays an important role only when $d\gg l$, in which one can separate the contributions from the fast-fast pairs and the other three kinds of entangled pairs, as studied in our current work. To demonstrate this, we checked the entanglement negativity evolution of two symmetric and disjoint intervals after a \emph{global} quench in a harmonic chain. It is found that in the case $d\gg l$, the numerical results are much smaller than the CFT results in magnitude, which agrees with our physical picture here.

\section{Discussions and conclusions}

\subsection{Relation between entanglement negativity and mutual information after a local quantum quench}

In the study of time evolution of the entanglement negativity $\mathcal{E}(t)$ after a global quench
\cite{Coser},
it is found that $\mathcal{E}(t)$ shows the same feature as that of the Renyi mutual information $I^{(n)}(t)$.
In our current study on the local quench problem, similar features are observed for the disjoint-interval cases.
In this part, we hope to understand this observation, from a more general point of view,
for both cases of adjacent and disjoint intervals
after a local quantum quench.
As a straightforward generalization of Ref.\ \cite{Coser},
it is found that the relation between the entanglement negativity and mutual information is independent of
whether the quench of our interest is global or local.
In other words, for both
global and local quenches,
the time evolution of the entanglement negativity $\mathcal{E}(t)$ has the same form as the Renyi mutual information $I^{n}(t)$ up to a global prefactor, as explicitly discussed in the following.

The mutual information between two adjacent/disjoint intervals $A_1$ and $A_2$ are defined as
\begin{equation}
I_{A_1,A_2}^{(n)}:=S_{A_1}^{(n)}+S_{A_2}^{(n)}-S_{A_1\cup A_2}^{(n)}.
\end{equation}
Based on the definition of Renyi entropy $S_{A}^{(n)}$ in Eq.\ (\ref{RenyiS}), we have
\begin{equation}
I_{A_1,A_2}^{(n)}:=\frac{1}{n-1}\ln \left(\frac{\text{Tr}\rho_{A_1\cup A_2}^n}{\text{Tr}\rho_{A_1}^n \text{Tr}\rho_{A_2}^n }\right).
\end{equation}
Let us
consider the disjoint-interval case first, \emph{i.e.}, $A_1=[z_{1},z_{2}]$ and $A_2=[z_3, z_4]$.
Expressed in terms of the correlation function of twist operators, we can get
\begin{equation}\label{RenyiT}
I_{A_1,A_2}^{(n)}=\frac{1}{n-1}\ln \left(\frac{\left\langle
\mathcal{T}_n(z_1)\bar{\mathcal{T}}_n(z_2)\mathcal{T}_n(z_3)\bar{\mathcal{T}}_n(z_4)
\right\rangle}{\left\langle
\mathcal{T}_n(z_1)\bar{\mathcal{T}}_n(z_2)\right\rangle\left\langle\mathcal{T}_n(z_3)\bar{\mathcal{T}}_n(z_4)
\right\rangle}\right).
\end{equation}
By using the conformal mapping in Eq.\ (\ref{conformalMap}), we have
\begin{equation}\label{RenyiT2}
\begin{split}
&\langle\prod_{i}\mathcal{T}_n(z_{2i-1})\bar{\mathcal{T}}_n(z_{2i})\rangle\\
=&\prod_{i}\left|\frac{dw}{dz}\right|_{z_i}^{\Delta_{(i)}}\langle\prod_{i}\mathcal{T}_n(w_{2i-1})\bar{\mathcal{T}}_n(w_{2i})\rangle_{\text{RHP}}
\end{split}
\end{equation}
where $\langle \cdots\rangle_{\text{RHP}}$ are correlators on the right half plane. The $2N$-point function of twist fields on the RHP reads
\begin{equation}
\begin{split}
&\langle\prod_{i}\mathcal{T}_n(w_{2i-1})\bar{\mathcal{T}}_n(w_{2i})\rangle_{\text{RHP}}\\
=&\frac{\tilde{c}_n^N}{\prod_{i}|w_i-\tilde{w}_i|^{\Delta_n}}\left(\frac{\prod_{j<k}\eta_{2k,2j}\eta_{2k-1,2j-1}}{\prod_{j,k}\eta_{2j-1,2k}}\right)^{\Delta_n}
\mathcal{F}(\{\eta_{j,k}\}),
\end{split}
\end{equation}
where the cross ratios $\eta_{i,j}$ are defined in Eq.\ (\ref{crossRatio}). Here we are only interested in the limit $\eta_{i,j}\to 0$ or $1$,
\emph{i.e.}, $\mathcal{F}(\{\eta_{j,k}\})$ is a constant. By neglecting various non-universal constants, then the Renyi mutual information in Eq.\ (\ref{RenyiT}) can be expressed as
\begin{equation}
I_{A_1,A_2}^{(n)}=\frac{1}{n-1}\ln\left(\frac{\eta_{1,3}\eta_{2,4}}{\eta_{1,4}\eta_{2,3}}\right)^{\Delta_n}.
\end{equation}
By using the expression of $\Delta_n$ in Eq.\ (\ref{ScalingDim}), one can immediately get
\begin{equation}
I_{A_1,A_2}^{(n)}=-\frac{c}{12}\cdot\frac{n+1}{n}\ln\left(\frac{\eta_{1,4}\eta_{2,3}}{\eta_{1,3}\eta_{2,4}}\right),
\end{equation}
which has the same form as $\mathcal{E}(t)$ in Eq.\ (\ref{ENdis}) apart from a constant prefactor $\frac{2}{3}$.
Then the Renyi mutual information
$I^{(n)}_{A_1\cup A_2}$ for two adjacent intervals can be obtained by taking the limit $z_3\to z_2$. Since
\begin{equation}\label{ME}
I^{(n)}_{A_1\cup A_2}(t)=\frac{2}{3}\frac{n+1}{n} \mathcal{E}_{A_1\cup A_2}(t)
\end{equation}
for two disjoint intervals, one can get
\begin{equation}
\lim_{z_3\to z_2} I^{(n)}_{A_1\cup A_2}(t)=\frac{2}{3}\frac{n+1}{n} \lim_{z_3\to z_2}\mathcal{E}_{A_1\cup A_2}(t).
\end{equation}

Although the above discussion is
for the local quench, one can find that the same conclusion holds for the global quench
\cite{Coser},
because our derivation is not sensitive to the concrete form of conformal mapping.

\subsection{Conclusions}
In this paper, we studied the time evolution of the entanglement negativity that results from a local quench in conformal field theories,
where the local quench is introduced by connecting two decoupled CFTs. Once the two CFTs are joint at the endpoints, the interaction is simultaneously introduced,
and then local quasiparticle excitations are generated at the jointing point. These quasiparticles carry information about entanglement and propagate freely along the system.
The entanglement negativity of two intervals are built through these propagating quasiparticles.

Then by employing CFT approach, we calculated the entanglement negativity evolution for both adjacent intervals and disjoint intervals respectively.
For the case of two adjacent intervals, the entanglement negativity grows as $\ln t$ initially,
and then develops a plateau-like feature.
When the quasiparticles propagate out of the intervals, the negativity drops to the ground-state value.
For the case of two disjoint intervals, there is no entanglement negativity at the very beginning of local quench until the quasiparticles
reach the two intervals simultaneously. In the limit $d\gg l$, a long-range entanglement which is independent of the distance $d$ between two intervals is built through the quasiparticles.
Then again, similar with the case of adjacent intervals, once the quasiparticles propagate out of the two intervals, there is a sharp drop of the entanglement negativity.

Because our results are obtained from the CFT approach, the conclusion only applies to critical systems with a linear dispersion relation in (1+1) dimensions.
In other words, in the CFT approach, all the quasiparticles propagate at the same speed (``speed of light'').
For a general lattice model such as the harmonic chain considered in this paper, however, the dispersion relation is not linear for all momentum vectors. There are some quasiparticles propagating slowly compared with the speed of light.
This is why the numerical results do not agree with the CFT results perfectly, although the main features agree with each other.

Last but not least, we mention some interesting future problems to be studied as follows.

$\bullet$ \emph{Finite temperature effects}.
Most recently, finite temperature effects on the entanglement negativity in conformal field theories
were studied
\cite{Calabrese1408}.
In particular,  finite temperature effects on a local quench of the entanglement negativity were studied numerically based on a critical harmonic chain
\cite{Viktor}.
It is interesting to generalize our method to the finite temperature case,
and have an analytical picture of the finite temperature effects on a local quench
of the negativity.

$\bullet$ \emph{Quantum entanglement of local operators, and entanglement density}.
In our work, a local quantum quench is realized by joining two separate CFTs at the endpoints.
Another method to realize a local quantum quench is through acting
with a local operator on an infinitely extended system
\cite{Tadashi1302, Tadashi1401,Tadashi1403}.
By studying the increase of the Renyi entropy at a later time,
quantum entanglement of local operators can be
detected.
In addition, by following the change of the entanglement entropy in a certain interval,
one can study the entanglement density,
which measures the density of entangled pairs between given two points
\cite{Tadashi1302,Tadashi1412}.
Here, in our work, the local quench of the negativity provides a natural platform
for studying the increase of entanglement,
and therefore may be used to extract the entanglement of local operators,
as well as the entanglement density.

$\bullet$ \emph{Check lattice models.}
In our work, to demonstrate the CFT results, we study the critical harmonic chain numerically.
It will be interesting to check the CFT predictions in more complicated lattice models,
such as itinerant fermions
\cite{Chang2015} and spin-chain systems, which are described by
the Luttinger liquid theory.

$\bullet$ \emph{Local quench of the entanglement negativity and entanglement renormalization.}
Entanglement renormalization provides a helpful framework to study the connection between quantum entanglement
and its holographic geometry
\cite{cMERA1102,cMERA1208,cMERA1311,cMERA1412}.
Global quenches and local quenches in entanglement renormalization are discussed in Refs.\ \cite{Tadashi1302,cMERA1311},
respectively.
In particular, the effect of local quench on the entanglement entropy evolution based on entanglement renormalization
is briefly discussed in Ref.\ \cite{Tadashi1302}.
It is of great interest to study the local quench of the entanglement negativity
within the framework of entanglement renormalization.

$\bullet$ \emph{Dynamical charged entanglement negativity}
Most recently, the concept of the charged Renyi entropies was proposed and studied in several works
\cite{Caputa,Shunji}.
In particular, in Ref.\ \cite{Caputa}, the dynamical evolution of the charged Renyi entropies was studied.
It is interesting to consider the entanglement negativity in the presence of angular momentum and $U(1)$ charge,
and study its dynamical properties after global quenches or local quenches.

$\bullet$ \emph{Holographic study of time evolution of entanglement negativity after a local quantum quench
}
It is worth mentioning that the mutual information after a local quench has been studied
with holographic approach recently
\cite{Tadashi1302, Asplund}.
On the other hand, holographic approach to the entanglement negativity was only discussed in the equilibrium case,
although more needs to be understood
\cite{Mukund}.
It will be of great interest to study the non-equilibrium properties of the entanglement negativity
with holographic methods. In addition, in a very recent work \cite{Hartman2015} on global quench, it is found that the free quasiparticle picture does not hold in a (1+1) dimensional CFT with $c>1$ (assuming no extended symmetry algebra).
Relevant features were also observed in Ref.\cite{Asplund}, where it is found that the holographic mutual information after a local quench shows qualitatively different features from the CFT results.
We believe that the same conclusion should hold in the local quench problem of the entanglement negativity, and we leave this for our future work.

\emph{Note added:} After we finished this work, we noticed that related results on local quenches of the entanglement negativity of two adjacent intervals appear in a recent work\cite{Hoogeveen}.
In Ref.\ \cite{Hoogeveen}, the time evolution of the entanglement negativity after a local quench was studied in finite temperature, and the authors mainly focus on the adjacent-interval case. We hope our current work provides a platform for the study of disjoint-interval case after a local quench in finite temperature.

\section{Acknowledgement}

X. W. thanks Yanxiang Shi for help with plotting. S. R. is supported by the Alfred P. Sloan foundation.
We thank the anonymous referees for useful comments.

\section{Appendix}
In the appendix, we give explicit forms of $w_{ij}$, which are used to calculate the entanglement negativity in the main text.
The conformal mapping we used is
\begin{equation}\label{conformalMapAppendix}
w=\frac{z}{\epsilon}+\sqrt{\left(\frac{z}{\epsilon}\right)^2+1}.
\end{equation}
where
\begin{equation}
z=l+i\tau.\nonumber
\end{equation}
Then one has
\begin{equation}
\begin{split}
w=&\frac{l+i\tau}{\epsilon}+\frac{1}{\epsilon}\sqrt{\epsilon^2+(l+i\tau)^2}\\
:=&\frac{1}{\epsilon}\left(l+i\tau+\rho e^{i\theta}\right)
\end{split}
\end{equation}
with
\begin{equation}
\left\{
\begin{split}
\rho=&\left[(\epsilon^2+l^2-\tau^2)^2+4l^2\tau^2
\right]^{1/4},\\
\theta=&\frac{1}{2}\text{arctan}\frac{2l\tau}{\epsilon^2+l^2-\tau^2}.
\end{split}
\right.
\end{equation}
The real time evolution can be obtained by replacing $\tau$ with $it$ in the last step. By expanding to the second order in $\epsilon$,
$\rho\cos\theta$ and $\rho\sin\theta$ are expressed as follows. For $l>0$, one has
\begin{small}
\begin{equation}\label{aprox}
\left\{
\begin{split}
\rho\cos\theta\to& \max[l,t]\left(1+\frac{\epsilon^2}{2\left[\left(\max[l,t]\right)^2-\left(\min[l,t]\right)^2\right]}\right)\\
\rho\sin\theta\to & i\min[l,t]\left(1+\frac{\epsilon^2}{2\left[\left(\min[l,t]\right)^2-\left(\max[l,t]\right)^2\right]}\right).\\
\end{split}
\right.\nonumber
\end{equation}
\end{small}
For $l<0$, one has
\begin{small}
\begin{equation}\label{aproxNegative}
\left\{
\begin{split}
\rho\cos\theta\to& \max[|l|,t]\left(1+\frac{\epsilon^2}{2\left[\left(\max[|l|,t]\right)^2-\left(\min[|l|,t]\right)^2\right]}\right)\\
\rho\sin\theta\to & -i\min[|l|,t]\left(1+\frac{\epsilon^2}{2\left[\left(\min[|l|,t]\right)^2-\left(\max[|l|,t]\right)^2\right]}\right).\\
\end{split}
\right.\nonumber
\end{equation}
\end{small}

\subsection{Appendix I: $w_{ij}$ for symmetric adjacent intervals}

\begin{equation}
w_{12}=\left\{
\begin{split}
&\sqrt{\frac{l}{l+t}}\ \ \ &t<l\\
&\frac{l}{\sqrt{t(t+l)}}, &t>l\\
\end{split}
\right.
\end{equation}
\begin{equation}
w_{1\tilde{2}}=\left\{
\begin{split}
&\sqrt{\frac{l}{l-t}}\ \ \ &t<l\\
&\frac{2}{\epsilon}\sqrt{t(t-l)}, &t>l\\
\end{split}
\right.
\end{equation}

\begin{equation}
w_{13}=\left\{
\begin{split}
&\frac{2}{\epsilon}\sqrt{l^2-t^2},\ \ \ &t<l\\
&2l\sqrt{\frac{1}{t^2-l^2}}, &t>l\\
\end{split}
\right.
\end{equation}

\begin{equation}
w_{1\tilde{3}}=\left\{
\begin{split}
&\frac{2}{\epsilon}\sqrt{l^2-t^2},\ \ \ &t<l\\
&\frac{2}{\epsilon}\sqrt{t^2-l^2}, &t>l\\
\end{split}
\right.
\end{equation}

\begin{equation}
w_{23}=\left\{
\begin{split}
&\frac{2}{\epsilon}\sqrt{l(l-t)},\ \ \ &t<l\\
&\sqrt{\frac{l^2}{t(t-l)}}, &t>l\\
\end{split}
\right.
\end{equation}

\begin{equation}
w_{2\tilde{3}}=\left\{
\begin{split}
&\frac{2}{\epsilon}\sqrt{l(l+t)},\ \ \ &t<l\\
&\frac{2}{\epsilon}\sqrt{t(l+t)}. &t>l\\
\end{split}
\right.
\end{equation}

Based on the expressions of $w_{ij}$, we can obtain the cross-ratio $\eta_{ij}$ as follows.
\begin{equation}
\eta_{12}=\left\{
\begin{split}
&\frac{l-t}{l+t},\ \ \ &t<l\\
&\frac{l^2\epsilon^2}{4}\cdot\frac{1}{t^2(t^2-l^2)}, &t>l\\
\end{split}
\right.
\end{equation}
\begin{equation}
\eta_{13}=\left\{
\begin{split}
&1,\ \ \ &t<l\\
&\frac{l^2\epsilon^2}{(t^2-l^2)^2}. &t>l\\
\end{split}
\right.
\end{equation}
\begin{equation}
\eta_{23}=\left\{
\begin{split}
&\frac{l-t}{l+t},\ \ \ &t<l\\
&\frac{l^2\epsilon^2}{4}\cdot\frac{1}{t^2(t^2-l^2)}. &t>l\\
\end{split}
\right.
\end{equation}
It is found that for the cases $t\ll l$, $t=l+0^-$ and $t>l$, one always has $\eta_{ij}=1$ or $0$.

\subsection{Appendix II: $w_{ij}$ for asymmetric adjacent intervals}

For asymmetric adjacent intervals, $w_{12}$ and $w_{1\tilde{2}}$ have the same form as those in the symmetric case, and we only need to change $l$ with $l_1$. Similarly, for $w_{23}$ and $w_{2\tilde{3}}$, we just change $l$ with $l_2$. Therefore, what we need to recalculate here are $w_{13}$ and $w_{1\tilde{3}}$ as follows. In the case of $l_1<l_2$, one has
\begin{equation}
w_{13}=\left\{
\begin{split}
&\frac{2}{\epsilon}\sqrt{l_2^2-t^2},\ \ \ &t<l_1\\
&\frac{2}{\epsilon}\sqrt{(l_1+l_2)(l_2-t)}, &l_1<t<l_2\\
&\sqrt{\frac{(l_1+l_2)^2}{(t-l_2)(t+l_1)}}, &t>l_2\\
\end{split}
\right.
\end{equation}
\begin{equation}
w_{1\tilde{3}}=\left\{
\begin{split}
&\frac{2}{\epsilon}\sqrt{l_2^2-t^2},\ \ \ &t<l_1\\
&\frac{2}{\epsilon}\sqrt{(t+l_2)(l_2-l_1)}, &l_1<t<l_2\\
&\frac{2}{\epsilon}\sqrt{(t+l_2)(t-l_1)}. &t>l_2\\
\end{split}
\right.
\end{equation}
In the case of $l_1>l_2$, one has
\begin{equation}
w_{13}=\left\{
\begin{split}
&\frac{2}{\epsilon}\sqrt{l_2^2-t^2},\ \ \ &t<l_2\\
&\sqrt{\frac{(l_1+l_2)(l_2+t)}{(t-l_2)(l_1+t)}}, &l_2<t<l_1\\
&\sqrt{\frac{(l_1+l_2)^2}{(t+l_1)(t-l_2)}}, &t>l_1\\
\end{split}
\right.
\end{equation}
\begin{equation}
w_{1\tilde{3}}=\left\{
\begin{split}
&\frac{2}{\epsilon}\sqrt{l_2^2-t^2},\ \ \ &t<l_2\\
&\sqrt{\frac{(l_1-l_2)(t+l_2)}{(l_1-t)(t-l_2)}}, &l_2<t<l_1\\
&\frac{2}{\epsilon}\sqrt{(t+l_2)(t-l_1)}. &t>l_1\\
\end{split}
\right.
\end{equation}

Without loss of generality, here we choose $l_1<l_2$, as what we did in the main text. Then one has
\begin{equation}
\eta_{12}=\left\{
\begin{split}
&\frac{l_1-t}{l_1+t},\ \ \ &t<l_1\\
&\frac{l_1^2\epsilon^2}{4}\cdot\frac{1}{t^2(t^2-l_1^2)}, &l_1<t<l_2\\
&\frac{l_1^2\epsilon^2}{4}\cdot\frac{1}{t^2(t^2-l_1^2)}, &t>l_2\\
\end{split}
\right.
\end{equation}
\begin{equation}
\eta_{23}=\left\{
\begin{split}
&\frac{l_2-t}{l_2+t},\ \ \ &t<l_1\\
&\frac{l_2-t}{l_2+t},\ \ \ &l_1<t<l_2\\
&\frac{l_2^2\epsilon^2}{4}\cdot\frac{1}{t^2(t^2-l_2^2)}. &t>l_2\\
\end{split}
\right.
\end{equation}
\begin{equation}
\eta_{13}=\left\{
\begin{split}
&1,\ \ \ &t<l_1\\
&\frac{(l_2-t)(l_1+l_2)}{(l_2+t)(l_2-l_1)},\ \ \ &l_1<t<l_2\\
&\frac{\epsilon^2(l_1+l_2)^2}{(t^2-l_2^2)(t^2-l_1^2)}, &t>l_2\\
\end{split}
\right.
\end{equation}
It is found that for the cases $t\ll l_1$, $t=l_2+0^-$ and $t>l_2$, one always has $\eta_{ij}=1$ or $0$.

\subsection{Appendix III: $w_{ij}$ for symmetric disjoint intervals}

\begin{equation}
w_{14}=\left\{
\begin{split}
&\frac{2}{\epsilon}\sqrt{(d+l+t)(d+l-t)},\ &t<d\\
&\frac{2}{\epsilon}\sqrt{(d+l+t)(d+l-t)}, &d<t<d+l\\
&2(d+l)\sqrt{\frac{1}{(t+d+l)(t-d-l)}}, & t>d+l
\end{split}
\right.
\end{equation}

\begin{equation}
w_{1\tilde{3}}=\left\{
\begin{split}
&\frac{2}{\epsilon}\sqrt{(d+t)(d-t)}, \ &t<d\\
&\sqrt{\frac{l(t+d)}{(t-d)(d+l-t)}}, &d<t<d+l\\
&\frac{2}{\epsilon}\sqrt{(t+d)(t-d-l)}, & t>d+l
\end{split}
\right.
\end{equation}

\begin{equation}
w_{23}=\left\{
\begin{split}
&\frac{2}{\epsilon}\sqrt{d^2-t^2},\ \ \ &t<d\\
&2d\sqrt{\frac{1}{t^2-d^2}}, &d<t<d+l\\
&2d\sqrt{\frac{1}{t^2-d^2}}, & t>d+l
\end{split}
\right.
\end{equation}

\begin{equation}
w_{2\tilde{4}}=\left\{
\begin{split}
&\frac{2}{\epsilon}\sqrt{(d+l+t)(d+l-t)},\ \ \ &t<d\\
&\frac{2}{\epsilon}\sqrt{l(d+l+t)}, &d<t<d+l\\
&\frac{2}{\epsilon}\sqrt{(d+l+t)(t-d)}, & t>d+l
\end{split}
\right.
\end{equation}

\begin{equation}\label{w51}
w_{1\tilde{4}}=\left\{
\begin{split}
&\frac{2}{\epsilon}\sqrt{(d+l+t)(d+l-t)},\ \ \ &t<d\\
&\frac{2}{\epsilon}\sqrt{(d+l+t)(d+l-t)}, &d<t<d+l\\
&\frac{2}{\epsilon}\sqrt{(t+d+l)(t-d-l)}, & t>d+l
\end{split}
\right.
\end{equation}

\begin{equation}\label{w31}
w_{13}=\left\{
\begin{split}
&\frac{2}{\epsilon}\sqrt{(d+t)(d-t)},\ \ \ &t<d\\
&\sqrt{(t+d)\left(\frac{1}{t-d}-\frac{1}{t+d+l}\right)}, &d<t<d+l\\
&\sqrt{(2d+l)\left(\frac{1}{t-d}-\frac{1}{t+d+l}\right)}, & t>d+l
\end{split}
\right.
\end{equation}

\begin{equation}
w_{2\tilde{3}}=\left\{
\begin{split}
&\frac{2}{\epsilon}\sqrt{d^2-t^2},\ \ \ &t<d\\
&\frac{2}{\epsilon}\sqrt{t^2-d^2}, &d<t<d+l\\
&\frac{2}{\epsilon}\sqrt{t^2-d^2}, & t>d+l
\end{split}
\right.
\end{equation}

\begin{equation}
w_{24}=\left\{
\begin{split}
&\frac{2}{\epsilon}\sqrt{(l+d+t)(l+d-t)},\ \ \ &t<d\\
&\frac{2}{\epsilon}\sqrt{(l+d-t)(l+2d)}, &d<t<d+l\\
&(2d+l)\sqrt{\frac{1}{(t+d)(t-d-l)}}. & t>d+l
\end{split}
\right.
\end{equation}

\subsection{Appendix IV: $w_{ij}$ for asymmetric disjoint intervals}

\begin{equation}
w_{13}=\left\{
\begin{split}
&\frac{2}{\epsilon}\sqrt{d_2^2-t^2},\ \ \ &t<d_1\\
&\frac{2}{\epsilon}\sqrt{d_2^2-t^2},\ \ \ &d_1<t<d_2\\
&\sqrt{\frac{(d_1+d_2+l)(d_2+t)}{(t-d_2)(d_1+l+t)}}, &d_2<t<d_1+l\\
&\sqrt{\frac{(d_1+d_2+l)^2}{(t+d_1+l)(t-d_2)}}, &d_1+l<t<d_2+l\\
&\sqrt{\frac{(d_1+d_2+l)^2}{(t+d_1+l)(t-d_2)}}, &t>d_2+l\\
\end{split}
\right.
\end{equation}

\begin{equation}
w_{1\tilde{3}}=\left\{
\begin{split}
&\frac{2}{\epsilon}\sqrt{d_2^2-t^2},\ \ \ &t<d_1\\
&\frac{2}{\epsilon}\sqrt{d_2^2-t^2},\ \ \ &d_1<t<d_2\\
&\sqrt{\frac{(d_1-d_2+l)(t+d_2)}{(d_1+l-t)(t-d_2)}},\ \ \ &d_2<t<d_1+l\\
&\frac{2}{\epsilon}\sqrt{(t+d_2)(t-d_1-l)}, &d_1+l<t<d_2+l\\
&\frac{2}{\epsilon}\sqrt{(t+d_2)(t-d_1-l)}, &t>d_2+l\\
\end{split}
\right.
\end{equation}

\begin{equation}
w_{14}=\left\{
\begin{split}
&\frac{2}{\epsilon}\sqrt{(d_2+l)^2-t^2},\ \ \ &t<d_1\\
&\frac{2}{\epsilon}\sqrt{(d_2+l)^2-t^2},\ \ \ &d_1<t<d_2\\
&\frac{2}{\epsilon}\sqrt{(d_2+l)^2-t^2},\ \ \ &d_2<t<d_1+l\\
&\frac{2}{\epsilon}\sqrt{(d_1+d_2+2l)(d_2+l-t)}, &d_1+l<t<d_2+l\\
&\sqrt{\frac{(d_1+d_2+2l)^2}{(t-d_2-l)(t+d_1+l)}}, &t>d_2+l\\
\end{split}
\right.
\end{equation}

\begin{equation}
w_{1\tilde{4}}=\left\{
\begin{split}
&\frac{2}{\epsilon}\sqrt{(d_2+l)^2-t^2},\ \ \ &t<d_1\\
&\frac{2}{\epsilon}\sqrt{(d_2+l)^2-t^2},\ \ \ &d_1<t<d_2\\
&\frac{2}{\epsilon}\sqrt{(d_2+l)^2-t^2},\ \ \ &d_2<t<d_1+l\\
&\frac{2}{\epsilon}\sqrt{(t+d_2+l)(d_2-d_1)}, &d_1+l<t<d_2+l\\
&\frac{2}{\epsilon}\sqrt{(t+d_2+l)(t-d_1-l)}, &t>d_2+l\\
\end{split}
\right.
\end{equation}

\begin{equation}
w_{23}=\left\{
\begin{split}
&\frac{2}{\epsilon}\sqrt{d_2^2-t^2},\ \ \ &t<d_1\\
&\frac{2}{\epsilon}\sqrt{(d_1+d_2)(d_2-t)}, &d_1<t<d_2\\
&\sqrt{\frac{(d_1+d_2)^2}{(t-d_2)(t+d_1)}}, &d_2<t<d_1+l\\
&\sqrt{\frac{(d_1+d_2)^2}{(t-d_2)(t+d_1)}}, &d_1+l<t<d_2+l\\
&\sqrt{\frac{(d_1+d_2)^2}{(t-d_2)(t+d_1)}}, &t>d_2+l\\
\end{split}
\right.
\end{equation}

\begin{equation}
w_{2\tilde{3}}=\left\{
\begin{split}
&\frac{2}{\epsilon}\sqrt{d_2^2-t^2},\ \ \ &t<d_1\\
&\frac{2}{\epsilon}\sqrt{(t+d_2)(d_2-d_1)}, &d_1<t<d_2\\
&\frac{2}{\epsilon}\sqrt{(t+d_2)(t-d_1)}, &d_2<t<d_1+l\\
&\frac{2}{\epsilon}\sqrt{(t+d_2)(t-d_1)}, &d_1+l<t<d_2+l\\
&\frac{2}{\epsilon}\sqrt{(t+d_2)(t-d_1)}, &t>d_2+l\\
\end{split}
\right.
\end{equation}

\begin{equation}
w_{24}=\left\{
\begin{split}
&\frac{2}{\epsilon}\sqrt{(d_2+l)^2-t^2},\ \ \ &t<d_1\\
&\frac{2}{\epsilon}\sqrt{(d_1+d_2+l)(d_2+l-t)}, &d_1<t<d_2\\
&\frac{2}{\epsilon}\sqrt{(d_1+d_2+l)(d_2+l-t)}, &d_2<t<d_1+l\\
&\frac{2}{\epsilon}\sqrt{(d_1+d_2+l)(d_2+l-t)}, &d_1+l<t<d_2+l\\
&\sqrt{\frac{(d_1+d_2+l)^2}{(t-d_2-l)(t+d_1)}}, &t>d_2+l\\
\end{split}
\right.
\end{equation}

\begin{equation}
w_{2\tilde{4}}=\left\{
\begin{split}
&\frac{2}{\epsilon}\sqrt{(d_2+l)^2-t^2},\ \ \ &t<d_1\\
&\frac{2}{\epsilon}\sqrt{(t+d_2+l)(d_2+l-d_1)}, &d_1<t<d_2\\
&\frac{2}{\epsilon}\sqrt{(t+d_2+l)(d_2+l-d_1)}, &d_2<t<d_1+l\\
&\frac{2}{\epsilon}\sqrt{(t+d_2+l)(d_2+l-d_1)}, &d_1+l<t<d_2+l\\
&\frac{2}{\epsilon}\sqrt{(t+d_2+l)(t-d_1)}. &t>d_2+l\\
\end{split}
\right.
\end{equation}

\bibliography{QuenchRef}

\end{document}